\documentclass[LectureNotes]{SciPost}  
\usepackage{amsmath,amssymb}
\usepackage{bm}
\usepackage{graphicx}
\usepackage{paralist}
\usepackage{url}
\usepackage{multirow}
\usepackage{diagbox}
\usepackage{footmisc}
\usepackage{tablefootnote}
\usepackage{hyperref}

\newcommand{\rd}{\ensuremath{\mathrm{d}}}
\newcommand{\bfd}{\ensuremath{\bm{d}}}
\newcommand{\bff}{\ensuremath{\bm{f}}}

\newcommand{\bfp}{\ensuremath{\bm{p}}}
\newcommand{\bfq}{\ensuremath{\bm{q}}}
\newcommand{\bfx}{\ensuremath{\bm{x}}}
\newcommand{\bfy}{\ensuremath{\bm{y}}}

\newcommand{\bftheta}{\ensuremath{\bm{\theta}}}
\newcommand{\bfphi}{\ensuremath{\bm{\phi}}}
\newcommand{\bfpsi}{\ensuremath{\bm{\psi}}}

\newcommand{\rmd}{\ensuremath{\text{d}}}

\usepackage{wasysym}
\newcommand{\stari}{{\star}} 
\newcommand{\starii}{{\ast}} 
\newcommand{\stariii}{{\hexstar}} 

\definecolor{RED}{rgb}{1,0,0}

\begin{document}

\begin{center}{\Large \textbf{
On Model Selection in Cosmology
}}\end{center}

\begin{center}
M.\,Kerscher\textsuperscript{1} and 
J.\,Weller\textsuperscript{1,2,3}
\end{center}

\begin{center}
  {\bf 1} Universit\"ats-Sternwarte
  Ludwig-Maximilians-Universit\"at M\"unchen,
  Scheinerstr.~1, 81679 M\"unchen, Germany\\
  {\bf 2} Max Planck Institute for Extraterrestrial Physics, 
  Giessenbachstrasse, 85748 Garching, Germany \\
  {\bf 3} Excellence Cluster Origins, 
  Boltzmannstr.~2, 85748 Garching, Germany \\[1.5em]
  martin.kerscher@lmu.de
\end{center}

\begin{center}
May 10, 2019
\end{center}

\section*{Abstract}
{\bf
  We review some of the common methods for model selection\footnote{The
    expression ``model selection'' sometimes ``model choice'' is well
    established.  In physics ``model comparison'' is presumably more
    appropriate, but we will stick with the de facto standard.}: the
  goodness of fit, the likelihood ratio test, Bayesian model selection
  using Bayes factors, and the classical as well as the Bayesian
  information theoretic approaches. 
  We illustrate these different approaches by comparing models for the
  expansion history of the Universe.
  In the discussion we highlight the premises and objectives entering
  these different approaches to model selection and finally recommend
  the information theoretic approach.  }

\tableofcontents

\section{Introduction}
\label{sec:intro}

\begin{figure}
  \centering
  \includegraphics[width=0.7\linewidth]{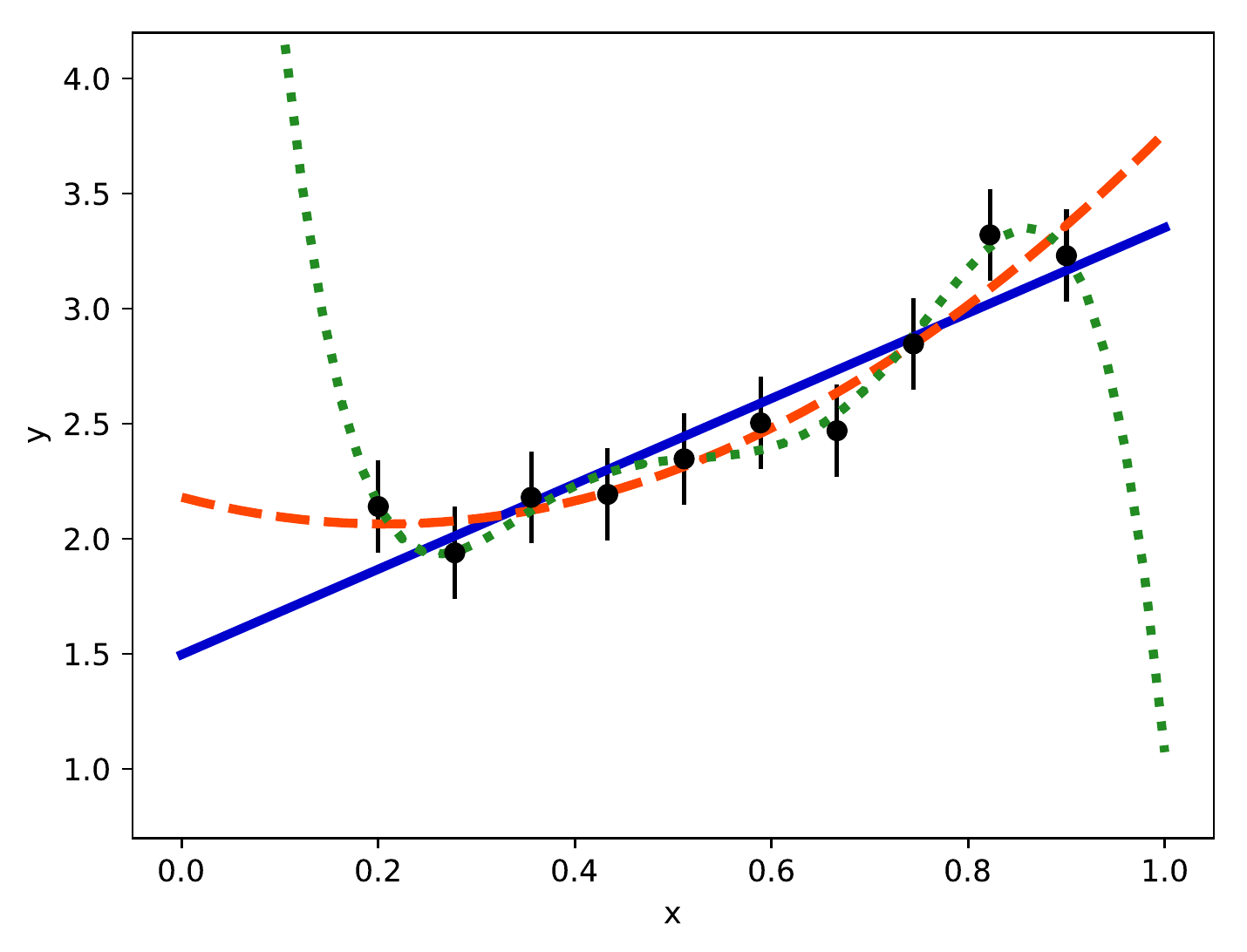}
  \caption{Three polynomials fitted to 10 data points:
  $f(x,\bftheta)= \theta_0 + \theta_1 x$ (solid blue line), 
  $g(x,\bfphi)=\phi_0+\phi_1 x+\phi_2 x^2$ (dashed red line)
  $h(x,\bfpsi)= \psi_0+\psi_1 x+\psi_2 x^2+\psi_3 x^3+\psi_4 x^4+\psi_5 x^5$ 
  (dotted green line). Which of these polynomials is the ``best'' model 
  (see also Munroe\,\cite{munroe:curvefitting})?
  }
  \label{fig:numquam}
\end{figure}

In science we often have several competing theoretical models which
try to explain the same natural phenomenon.  Based on measured data we
want to decide which model is the better one.
As an example we consider two different cosmological models, a cold
dark matter model (CDM) with a cosmological constant $\Lambda$ called
$\Lambda$CDM, and a cold dark matter model with a constant equation of
state $p=w\varrho$ for the dark energy component called $wCDM$.  With
both models we try to explain observations like the cosmic microwave
background, galaxy cluster counts, supernovae distance measurements,
to name only a few.
Models with more parameters typically allow for a closer fit of the
data, but are such models with more parameters indeed better (see
Fig.~\ref{fig:numquam})?  In such a context one often refers to
Ockham's razor that one should not introduce additional parameters if
they are not needed\footnote{Numquam ponenda est pluralitas sine
  necessitate. Attributed to William of Ockham and sometimes earlier
  to Duns Scotus. See Thorburn\,\cite{thorburn:myth} for an historical
  account.}.  One task of model selection is to make this statement
quantitative.

To set a well defined stage, consider some measured data points
$d_i=(x_i,y_i)$ with $i\in\{1,\ldots,N\}$.  A model is providing a
function $f(x,\bftheta)$ such that $f(x_i,\bftheta)$ is approximating
$y_i$ for each $i$.  The parameters
$\bftheta=(\theta_1,\ldots,\theta_K)$ are from $A\subset\mathbb{R}^K$.
For simplicity we assume that $x,x_i,y_i\in\mathbb{R}$ and also
$f(x,\bftheta)$ is real valued\footnote{Choosing a real valued $f$ and
  real $x,x_i,y_i$ is done for notational simplicity. We could choose
  a more complex mapping without touching the following arguments. }.
The best fitting parameters $\bftheta^\stari\in A\subset\mathbb{R}^K$
of the model are then determined from a Bayesian approach, a maximum
likelihood procedure, or a simple least--square--fit.
If one considers only a single model and has a good idea about the
priors for the parameters, then many physicists would agree that a
Bayesian parameter estimation procedure is the appropriate thing to do
(see \cite{cousins:why}).

The situation is more complicated if one considers at least one other
model $g(x,\bfphi)$ with parameters $\bfphi\in B\subset\mathbb{R}^L$.
For simplicity we name the models after the functions $f$ and $g$.
Typically the dimensions of the parameter spaces differ $L\ne K$ and
also the parameter spaces may not overlap.  We can determine the
optimal parameters $\bftheta^\stari$ and $\bfphi^\stari$ for each of
the models $f$ and $g$. But the question remains, which of the models
is ``better''.  This situation is illustrated in
Fig.~\ref{fig:numquam}.

As a starting point we will briefly discuss some of the common methods
used for parameter estimation.  Then we will present methods used for
the selection of models and also comment on the approximations and
numerical methods used.
In section\,\ref{sec:application} we use these methods to compare
two models for the expansion history of the Universe.
In the discussion we highlight the premises and objectives entering
the different approaches and recommend the information theoretic 
procedure, preferably in its Bayesian flavour.
In appendix~\ref{sect:stat} we summarise properties of statistical
tests, the empirical distribution function, and the Kullback-Leibler
divergence. In appendix~\ref{sec:implement} we provide some details of
the numerical implementation and in appendix~\ref{sec:errors} we
discuss the application and especially the error budget in more
detail.
We assume some familiarity with statistical methods used in physics
and cosmology (see for example \cite[chap.~40]{patrignani:pdg} for a
short review).
A comprehensive introduction to statistics including model selection
is Wassermann\cite{wassermann:all}. An introduction to model selection
with a focus on the information theoretic approach is Burnham \&
Anderson\cite{burnham:model}.
Most of the material shown here is not new.  Some of the results are
scattered throughout the literature so we try to present them in a
coherent fashion and give due reference.

\subsection{Parameter estimation}

In the following we will give a short review of different methods for
parameter estimation.  The starting point for a comparison of a model
with the data is in most cases the likelihood. To specify the
likelihood function $p_f(\bfd\,|\,\bftheta)$, the model itself
$f(x,\bftheta)$ and an error model for the data is needed. The vector
of measurements is $\bfd=(x_i, y_i)_{i=1}^N$ and
$p_f(\bfd\,|\,\bftheta)$ is the probability of obtaining the measured
data points $\bfd$, given the parameters $\bftheta$ in the model $f$.
Often one assumes Gaussian errors, then the likelihood reads
\begin{equation}
\label{eq:gauss-likelihood}
p_f(\bfd\,|\,\bftheta) = 
\frac{1}{\left((2 \pi )^N \det(\Sigma)\right)^{\frac{1}{2}}} \ 
\exp\left[- \tfrac{1}{2}
  (\bfy-\bff(\bfx,\bftheta))^T \Sigma^{-1} (\bfy-\bff(\bfx,\bftheta)) \right]  
\end{equation}
with $\bfy=(y_1,\ldots,y_N)^T$, $\bfx=(x_1,\ldots,x_N)^T$,
$\bff(\bfx,\bftheta)=(f(x_1,\bftheta),\ldots,f(x_N,\bftheta))^T$ and
the covariance matrix $\Sigma$.
With a maximum likelihood estimator we determine the parameters
$\bftheta^\stari$ which are maximising
$p_f(\bfd\,|\,\bftheta^\stari)$. Hence in choosing the parameter
$\bftheta^\stari$, the data points $\bfd$ become the most probable
data points given the model $f$.

The least square method is a simplified version of the maximum
likelihood estimator \cite{cramer:elements}.  The likelihood is
assumed to be Gaussian with a diagonal $\Sigma$ and the $\sigma_i$'s
on the diagonal.  Searching the maximum of $p_f(\bfd\,|\,\bftheta)$,
or of
\begin{equation}
\log p_f(\bfd\,|\,\bftheta) =
-\tfrac{N}{2}\log(2\pi) - \tfrac{1}{2}\log( \det(\Sigma))
- \tfrac{1}{2} \sum_{i=1}^N \frac{(y_i -  f(x_i,\bftheta) )^2}{\sigma_i^2}
\end{equation}
gives the same result as searching for the minimum of
\begin{equation}
\label{eq:chi2}
\chi^2_f = \sum_{i=1}^N \frac{(y_i -  f(x_i,\bftheta) )^2}{\sigma_i^2} .
\end{equation}
This minimum determines the best parameters $\bftheta^\starii$ of the
least square fit.

In a Bayesian setting we also need the prior distribution
$p_f(\bftheta)$ of the parameters for model $f$.  Using Bayes theorem
we can determine the posterior distribution
\begin{equation}
\label{eq:posterior}
p_f(\bftheta\,|\,\bfd) 
= \frac{p_f(\bfd\,|\,\bftheta)\ p_f(\bftheta)}{p_f(\bfd)},
\end{equation}
the distribution of the parameters $\bftheta$, given the data $\bfd$
and the model $f$.  Contrary to the posterior distribution
$p_f(\bftheta\,|\,\bfd)$, the likelihood $p_f(\bfd\,|\,\bftheta)$ is
the distribution of the data $\bfd$ given the parameters $\bftheta$ of
the model $f$.
The normalisation $p_f(\bfd)$ is called the evidence or marginal
likelihood.  The evidence can be obtained by an integration in
parameter space
\begin{equation}
\label{eq:evidence}
p_f(\bfd) 
= \int p_f(\bfd\,|\,\bftheta)\ p_f(\bftheta) \mathrm{d}\bftheta .
\end{equation}
Given the data and the model, the evidence is a normalisation constant
of the posterior distribution.  Hence, we do not need to calculate the
evidence if we want to determine the maximum $\bftheta^\stariii$ or
the mean of the posterior distribution
$p_f(\bftheta^\stariii\,|\,\bfd)$ only.  The maximum\footnote{The
  different ``best'' parameter estimates receive different stars as
  labels: $\bftheta^\starii$ (least--square) $\bftheta^\stari$
  (maximum--likelihood), $\bftheta^\stariii$ (MAP).}
$\bftheta^\stariii$ is called the maximum posterior (MAP) estimate.
Clearly, there is more to parameter estimation than we covered
here. We did not discuss how to choose priors, how to deal with
nuisance parameters, or how to determine confidence or credibility
regions. Nevertheless we provided the necessary prerequisites to be
able to discuss model selection.

\section{Model selection}
\label{sec:modelcompare}

Quite a few methods for model selection have been developed.
Cosmological and astrophysical oriented reviews and books are for
example \cite{hobson:bayesian}, \cite{verde:statistical}
\cite{lupton:statistics}. A more philosophically inclined introduction
with basic examples can be found in
Sober\,\cite[chapter~1]{sober:evidence}.

\subsection{The goodness of fit}
\label{sect:goodness}

The so called ``goodness of fit'' may serve as a starting point for
this discussion, since it is often the first method students of
physics learn in their lab--courses.
One calculates the so called reduced $\chi^2_\text{f,red}$
\begin{equation}
\chi^2_{f,\text{red}} = \frac{\chi^2_f}{n_\text{df}}
\end{equation}
where $\chi^2_f$ is calculated using
eq.\,(\ref{eq:chi2}) with the best fit parameters $\bftheta^\starii$
of the model.  $n_\text{df}$ is the number of degrees of
freedom, typically $n_\text{df}=N-K$, with $N$ the number of data
points and $K$ the number of parameters in the model $f$.
If $\chi^2_{f,\text{red}}\approx1$, this is considered a good fit, 
if $\chi^2_{f,\text{red}}>1$ a bad fit and 
if $\chi^2_{f,\text{red}}<1$ an overfit.
Remember, one starts with a maximum likelihood estimate and makes
assumptions about the data- and error-model which are often stark
oversimplifications.
As a remedy the number of degrees of freedom is determined from the
``effective'' number of independent data points.  The problems of this
approach are summarised in \cite{andrae:dosanddonts}.

Some of the motivation for using the $\chi^2_f$ derives from the
theory of statistical tests (see \cite{kendall:advanced2} or
\cite{lupton:statistics}). The $p$--value used in these tests is 
calculated as 
\begin{equation}
p = 1 - G_{n_\text{df}}(\chi^2_f),
\end{equation}
with $G_{n_\text{df}}$ the cumulative probability distribution
function of a $\chi^2$ distributed random variable with
${n_\text{df}}$ degrees of freedom (see appendix\,\ref{sect:stat}). 
Clearly $p$ is one-to-one with $\chi^2_f$.
The $p$--value indicates how incompatible the data are with our
null~hypothesis (our model including the error model,
\cite{wasserstein:asa}).  The smaller the $p$--value, the greater is
the statistical incompatibility of the data with the null~hypothesis.
Hence, given the model (the null~hypothesis), a small $p$--value
allows us to reject the model using a statistical test.
From the $p$--value however we do not learn anything about the false
negative rate (see appendix\,\ref{sect:stat}).  The $p$--value is a
statement about data in relation to a specified hypothetical
explanation (our model), not about the data itself, and especially not
about other models. See the recent statement of the American
Statistical Association on the (restricted) applicability of
$p$--values \cite{wasserstein:asa}.

Whether an alternative hypothesis/model is needed, was one of the
issues in the debate about hypothesis testing between Fisher on one
side and Neyman and Pearson on the other side (see
Lehmann~\cite{lehmann:fisher} for a breakdown of the arguments). With
respect to the importance of the false negative rate and the
specification of an alternative hypothesis we side here with Neymann
and Pearson (see the following section\,\ref{sec:lrt}).

\subsection{Likelihood ratio test}
\label{sec:lrt}

For the selection of models Neyman and Pearson\,\cite{neyman:problem}
developed the likelihood ratio test.  As usual we first consider so
called nested models.  The model $f$ with parameter space $A$ is a
special case of the model $g$ with parameter space $B$. More formally
$A\subsetneq B$ and $f\vert_A \equiv g\vert_A$ restricted to $A$.  Now
we determine the best fitting parameter $\bftheta^\stari\in A$, and
$\bfphi^\stari\in B$ and calculate
\begin{equation}
L = \frac{p_f(\bfd\,|\,\bftheta^\stari)}{p_g(\bfd\,|\,\bfphi^\stari)} .
\end{equation}
Our null~hypothesis is ``$f$ is the true model with
$\bftheta^\stari\in A$''.  The alternative is ``$g$ is the true model
with $\bfphi^\stari\in B$ but $\bfphi^\stari\not\in A$''.  With these
maximum likelihood estimates $\bftheta^\stari$, and $\bfphi^\stari$ we
calculate $L$. Fixing a significance level $0<\alpha<1$ one can
proceed and specify the test. Often one relies on Wilk's theorem
\cite{wilks:largesample}: for nested models and for large sample sizes
$N$ the $\lambda = -2\log L$ is approximately $\chi^2$--distributed
with the number of degrees of freedom equal to
$\nu=\text{dim}(B)-\text{dim}(A)$. The $p$--value is calculated as
$p=1-G_\nu(\lambda)$. We reject our null hypothesis if $p<\alpha$ with
a predefined significance level $\alpha$ (see the
appendix~\ref{sect:stat}).
Contrary to the situation discussed with the goodness of fit, the
alternative hypothesis is fully specified in the likelihood ratio
test.  The false negative rate\footnote{The false negative rate is
  also called the type~II or $\beta$ error.} is the probability that
the true alternative hypothesis (our model g) gets rejected.
The Neyman-Pearson--Lemma\cite{neyman:problem} tells us that a test
based on the likelihood ratio is minimising the false negative rate.
In this sense the likelihood ratio test is optimal.

In the introduction we already considered a more general, non-nested
setting. Vuong\cite{vuong:likelihood} discusses the likelihood ratio
test for overlapping or non-nested models and he derives the relevant
limiting distribution (not necessarily a $\chi^2$-distribution
anymore). The application of the likelihood ratio test in this more
general setting is reviewed by Lewis et al.\cite{lewis:unified}.

\subsection{Bayesian model selection}
\label{sect:bayes}

Bayesian methods, like the evidence and Bayes factors are nowadays
frequently used to compare cosmological models (see for example
\cite{saini:revealing}, \cite{trotta:sky},
\cite{tang:complementarity}, and \cite{kitching:model}).
The definition of the evidence in eq.\,(\ref{eq:evidence}) 
\[
p_f(\bfd) 
= \int p_f(\bfd\,|\,\bftheta)\ p_f(\bftheta) \mathrm{d}\bftheta 
\]
tells us that $p_f(\bfd)$ is the conditional probability of obtaining
the data vector $\bfd$ given the model $f$.
For some simple models the evidence can be calculated and a suggestive
interpretation emerges\cite{saini:revealing}, but in most cases the
evidence of one model by itself is not very informative. Its
usefulness derives from the evidence ratio used in Bayesian model
selection.

If we consider another model $g$ we may compare its evidence with the
evidence of the model $f$.  For a consistent comparison of models
within a Bayesian framework we need the joint probability
$p(f\,\text{and}\,\bfd) = p_f(\bfd) \pi_f$ of model $f$ and data
$\bfd$. Similarly for $p(g\,\text{and}\,\bfd) = p_g(\bfd) \pi_g$ of
model $g$ and the same data $\bfd$. The $\pi_f$ and $\pi_g$ are the
prior probabilities we assign to our models.  Often these
probabilities are chosen equal $\pi_f=\pi_g$, and the ratio of the
full probabilities
\begin{equation}
  \frac{p(f\,\text{and}\, \bfd)}{p(g\,\text{and}\,\bfd)} 
  = \frac{p_f(\bfd)\,\pi_f}{p_g(\bfd)\,\pi_g}
  = \frac{p_f(\bfd)}{p_g(\bfd)} =: B_{fg}
\end{equation}
reduces to the evidence ratio, also called Bayes factor.  A $B_{fg}$
larger than unity suggests, that we should favour model\footnote{We
  will comment on Jeffreys' scale in section\,\ref{sec:application};
  see also footnote \ref{foot:jeffreys}.}  $f$ over model $g$.

The Bayes factor, as any result from a Bayesian analysis, explicitly
depends on the prior distributions for the parameters of the models.  
You may have prior knowledge that allows you to specify a so called
``subjective'' prior \cite{goldstein:subjective}. Practitioners often
use priors suggested by results from preceding observations or
studies.
Different approaches are used to motivate the so called ``objective'',
``non-informative'', or ``reference'' priors. Their definition can be
based on the principle of insufficient reasoning, the maximum entropy
principle, the invariance under transformations or scaling, or the
missing information principle (see e.g.\ \cite{jeffreys:theory},
\cite{jaynes:prior}, \cite{bernardo:reference}).  Kass\,\&\,Wassermann
\cite{kass:selection} provide an overview and rules for selecting
among these priors. For a stimulating dialogue with J.M.\,Bernardo on
prior probabilities see \cite{irony:noninformative} (don't miss the
comments on this dialogue by D.R.\,Cox, A.P.\,Dawid, J.K.\,Ghosh and
D.\,Lindley in the same issue).  In any case, it is important to
select the prior carefully, and it seems advisable to investigate the
dependency of the model selection on the prior.

The calculation of the evidence (eq.\,(\ref{eq:evidence})) can be
quite challenging. Friel\,\&\,Wyse~\cite{friel:estimating} provide a
review of different techniques.
One of the first approximations for the evidence is due to
Schwarz\,\cite{schwarz:estimating}. Asymptotically he arrives at the
so called Bayesian Information Criterium\footnote{This name Bayesian
  \emph{information} criterium is unfortunate, no information theory
  is involved here. Burnham \& Anderson\cite{burnham:multimodel} argue
  that the information theoretically motivated AIC (see next section)
  is a Bayesian procedure with a special prior.}  (see
\cite{neath:bic} for a detailed derivation)
\begin{equation}
\label{eq:BIC}
\text{BIC}(f) 
= -2\, \sum_{i=1}^N \log\,p_f(d_i\,|\,\bftheta^\stariii) + K \log\,N.
\end{equation}
If we compare models, a smaller value of the BIC is better.
%
The marginalised likelihood $p_f(d_i\,|\,\bftheta^\stariii)$
used in eq.\,(\ref{eq:BIC}) is obtained by fixing
$d_i=(x_{i},y_{i})$
and integrating over the remaining
$\bfd_{[i]}=((x_1,y_1),\ldots,(x_{i-1},y_{i-1}),(x_{i+1},y_{i+1}),\ldots,(x_N,y_N))^T$,
\begin{equation}
\label{eq:marginallikelihood}
p_f(d_i\,|\,\bftheta) = \int p_f(\bfd\,|\,\bftheta)\, \rd \bfd_{[i]} .
\end{equation} 
For a Gaussian likelihood with covariance matrix $\Sigma$ as in
eq.\,(\ref{eq:gauss-likelihood}), the integration can be readily
performed and the marginalised likelihood is a one dimensional
Gaussian with variance $\Sigma_{ii}$.

Beyond this asymptotic approach, several numerical techniques are
currently used to calculate the evidence.  In low dimensional
parameter spaces a direct integration using standard numerical methods
is sometimes possible.
In cosmology a method derived from the ideas of
Chib\,\cite{chib:marginal} has been used to estimate the evidence from
a given MCMC chain \cite{heavens:marginal,heavens:noevidence}.
With nested sampling one estimates the evidence directly
\cite{skilling:nested}.  Several implementations of this approach are
currently in use (see e.g.\ \cite{brewer:dnest4} and
\cite{feroz:importance} and references therein).
Kilbinger et al.\,\cite{kilbinger:bayesian} suggest a population Monte
Carlo method to calculate the evidence.
Another approach to estimate the Bayes factor is via the Savage-Dickey
density ratio \cite{verdinelli:computing}.
Comparisons of further numerical methods are discussed by
\cite{bos:comparison}, \cite{ardia:comparative}, and
\cite{friel:estimating}.

\subsection{Information theoretic approach to model selection}

The information theoretic approach is based on the concept of
minimising the distance between the distribution of the model and the
distribution of the data.
We assume that some observational data $d$ is drawn at random from the
true but unknown distribution with probability density $p_T(d)$.  From
our model $f$ and the data $\bfd=\{d_i\}_{i=1}^N$ we construct a
predictive distribution $p_{p,f}(d)$ for a single new observation $d$.
Several possibilities for such a predictive distribution exist and we
will discuss the classical and the Bayesian approach.
For now we assume that we know such a predictive distribution
$p_{p,f}(d)$ for our model $f$ which we want to compare to the true
distribution $p_T(d)$.  We measure the discrepancy between the two
distributions using the Kullback-Leibler (KL) divergence (see the
appendix~\ref{sect:stat})
\begin{align}
  D(p_T\,|\,p_{p,f})
  & = \int p_T(d) \log\frac{p_T(d)}{p_{p,f}(d)}\, \rd d  \nonumber \\
  & = \int p_T(d) \log p_T(d)\,\rd d  - \int p_T(d) \log p_{p,f}(d)\,\rd d .
\label{eq:KL-model-truth}
\end{align}
For model selection we rank the models $f$ and $g$ according to the
value of $D(p_T\,|\,p_{p,f})$ and $D(p_T\,|\,p_{p,g})$ --- the smaller
the better.

\subsubsection{Classical information theoretic approach}
\label{sec:information-classic}

In the classical information theoretic approach to model selection the
predictive likelihood is used as the predictive distribution.  This
leads to the so called Akaike Information Criterion (AIC, see
\cite{akaike:information,akaike:likelihood}).  Applications of the AIC
in cosmology are discussed in \cite{liddle:information} and
\cite{szydlowski:aic}.
Although several definitions of a predictive likelihood exist (see
\cite{bjornstad:predictive} for a review) we follow
\cite{akaike:likelihood} and use the marginalised likelihood
eq.\,(\ref{eq:marginallikelihood}) as the predictive likelihood
$p_{p,f}(d) \equiv p_{f}(d\,|\,\bftheta^\stari)$. This is the
likelihood of a new data point $d$ assuming the maximum likelihood
estimate $\bftheta^\stari$ of the parameters.
This is already a well defined approach if $d$ is a simple random
variable. However in our regression setting we have $d=(x, y)$ and we
compare the predictions of the model $f(x, \bftheta)$ to the observed
value $y$. For each of the observed data points $d_i=(x_i, y_i)$ we
know the uncertainties of the measurements entering the likelihood
(compare eq.\,(\ref{eq:gauss-likelihood}) and
eq.\,(\ref{eq:marginallikelihood})). But how do we calculate
$p_{f}(d\,|\,\bftheta^\stari)$ for a $d\ne d_i$?  At a first glance it
seems necessary to introduce an additional model for the
uncertainties.  As an example we could interpolate between
neighbouring values of the marginalised likelihood
(eq.\,(\ref{eq:marginallikelihood})) to determine
$p_{f}(d\,|\,\bftheta^\stari)$.  Fortunately we will see below, that
this is not necessary since we evaluate $p_{f}(d\,|\,\bftheta)$ only
at $d=d_i=(x_i,y_i)$.

Let us start with the derivation of the AIC (following loosely
\cite{konishi:generalised}).  The first term on the second line in
eq.\,(\ref{eq:KL-model-truth}) does not depend on the model $f$. The
second term is the expected log likelihood for the model $f$ for all
possible data
\begin{equation}
\eta(f)  = \int \log p_{p,f}(d)\, p_T(d)\rd d 
 = \int \log p_f(d\,|\,\bftheta^\stari)\, \rd F_T(d).
\end{equation}
This expectation is calculated using the true cumulative distribution
$F_T$.  Unfortunately the true cumulative distribution $F_T$ is
unknown. From the observational data $\bfd=(x_i,y_i)_{i=1}^N$ it is
always possible to construct the empirical cumulative distribution
function $F_{T,N}(d)$ as a sum of step functions (see
appendix~\ref{sect:stat}). Then an estimate of the expected log
likelihood $\eta(f)$ is given by
\begin{align}
  \label{eq:etahat}
  \widehat{\eta}(f) 
   = \int \log p_f(d\,|\,\bftheta^\stari)\, \rd F_{T,N}(d) 
   = \frac{1}{N} \sum_{i=1}^N \log p_f(d_i\,|\,\bftheta^\stari) ,
\end{align}
where we used
$\rd F_{T,N}(d)= \frac{1}{N} \sum_{i=1}^N \delta_{d_i}(d) \rd d$
(compare eq.\,(\ref{eq:empiricaldist})).
Since we estimated the best parameter $\bftheta^\stari$ from the same
dataset we use to construct the empirical distribution function
$F_{T,N}(d)$, the $\widehat{\eta}(f)$ is a biased estimate of
$\eta(f)$.  The expected bias of $\widehat{\eta}(f) $ is
\begin{equation}
  \label{eq:bf}
  b(f) = \int  \left( \widehat{\eta}(f) - \eta(f)\right) \, \rd F_T
\end{equation}
and the bias corrected expected log likelihood (the second term 
in  eq.\,(\ref{eq:KL-model-truth})) is 
\begin{equation}
  \label{eq:correction}
  \widehat{\eta}(f) - b(f) .
\end{equation}
Clearly, this only shifts the problem from $\widehat{\eta}(f)$ to
$b(f)$. 
Assuming that the true distribution $p_T$ is part of the family of
distributions $p_f(z\,|\,\bftheta)$ and that $\bftheta^\stari$ is a
maximum likelihood estimate, Akaike\cite{akaike:information} shows
that $b(f)$ asymptotically has the form $K/N$, with $K$ the dimension
of the parameter space and $N$ the number of data points.  Rescaling
this approximate expression of eq.\,(\ref{eq:correction}) by $-2N$ we
arrive at the Akaike Information Criterium\footnote{We follow the
  convention used by H.\,Akaike \cite{akaike:information}.}
\begin{equation}
\label{eq:AIC}
\text{AIC}(f) = -2N \left( \widehat{\eta}(f) - K/N\right)
= -2\,\sum_{i=1}^N \log p_f(d_i\,|\,\bftheta^\stari) + 2\,K.
\end{equation}
The model with the smaller value of the AIC is favoured. 
In Appendix\,\ref{sec:bootstrap} we detail the bootstrap method of
Konishi\,\&\,Kitagawa \cite{konishi:generalised} to obtain an estimate
$\widetilde{b}(f)$ for the bias $b(f)$. This allows the definition of
the Extended Information Criterium \cite{ishiguro:bootstrapping,
  konishi:generalised}
\begin{equation}
\label{eq:EIC}
\text{EIC}(f) = -2N \left( \widehat{\eta}(f) - \widetilde{b}(f) \right).
\end{equation}
The model with the smaller value of the EIC is favoured.

A comparison of eq.\,(\ref{eq:AIC}) with eq.\,(\ref{eq:BIC}) shows
that the AIC and the BIC differ in how they disfavour high dimensional
parameter spaces. These terms are sometimes called Ockham's razor
terms.  Keep in mind that the derivations of the AIC and the BIC start
from different principles: the BIC starts from the evidence and the
AIC from the proximity of a model to the true distribution.
Several extensions and ``corrections'' to the AIC have been proposed
(see for example \cite{stone:asymptotic}).  A corrected AIC, better
suited for smaller sample sizes, was derived by
\cite{hurvich:regression} (see \cite{cavanaugh:unifying} for a
unifying derivation of AIC and AICc).
\begin{equation}
\text{AICc}(f) = 
-2\sum_{i=1}^N \log p_f(d_i\,|\,\bftheta^\stari) + 2K\,\frac{N-K-1}{N}.
\end{equation}
Several of the assumptions, entering the derivation of the AIC, can be
relaxed and the estimates of the bias $b(f)$ can be improved (see
\cite{konishi:generalised} for a summary).
Indeed $\bftheta^\stari$ need not be a maximum likelihood estimate;
the asymptotic of $b(f)$ is known for Fisher consistent estimates
$\bftheta^\stari$ and also for MAP estimates $\bftheta^\stariii$, as
obtained from a Bayesian parameter estimation procedure.
However we are only aware of the bootstrap procedure discussed by
\cite{ishiguro:bootstrapping, konishi:generalised} as a direct
numerical approach to estimate $\eta(f)$ (see
appendix~\ref{sec:bootstrap}).

\subsubsection{Bayesian information theoretic approach}
\label{sec:bayesinfo}

In the classical approach we use the best fit marginalised likelihood
$p_f(d\,|\,\bftheta^\stari)$ as the predictive distribution. In a
Bayesian approach we use the posterior predictive distribution
$p_{p,f}(d)\equiv p_{\text{ppd},f}(d)$.
With the posterior distribution $p_f(\bftheta\,|\,\bfd)$ for the
parameters given in eq.\,(\ref{eq:posterior}) and the marginalised
likelihood $p_f(d\,|\,\bftheta)$ from
eq.\,(\ref{eq:marginallikelihood}) we can define the posterior
predictive distribution
\begin{equation}
  p_{\text{ppd},f}(d) 
  = \int p_f(d\,|\,\bftheta)\,  p_f(\bftheta\,|\,\bfd)\, \rd \bftheta .
\end{equation}
With $p_{\text{ppd},f}(d)$ in eq.\,(\ref{eq:KL-model-truth}) we
compare the posterior predictive distribution to the true distribution
$p_T(\bfd)$ using the KL-divergence:
\begin{multline}
  \label{eq:bayes_kl}
  D(p_T\,|\,p_{\text{ppd},f})
   = \int p_T(d) \log p_T(d)\,\rd d  -
   \int p_T(d)\, \log\left(  \int p_f(d\,|\,\bftheta)\, 
   p_f(\bftheta\,|\,\bfd)\, \rd \bftheta \right) \rd d . 
\end{multline}
The first term does not depend on the model $f$ and the last term in
eq.\,(\ref{eq:bayes_kl}) is
\begin{equation}
\kappa(f)  = \int 
  \log\left( \int p_f(d\,|\,\bftheta)\, p_f(\bftheta\,|\,\bfd)\, 
    \rd \bftheta\right)\ 
 \rd F_T(d).
\end{equation}
We follow the strategy from Sect.\,\ref{sec:information-classic} and
insert the empirical distribution function $F_{T,N}(d)$ for $F_T(d)$
and obtain
\begin{equation}
  \label{eq:hatkappa}
  \widehat{\kappa}(f) = \frac{1}{N} \sum_{i=1}^N \log\left(
  \int p_f(d_i\,|\,\bftheta)\, p_f(\bftheta\,|\,\bfd)\,\rd\bftheta \right)
  = \frac{1}{N} \sum_{i=1}^N 
  \log\left( \mathbb{E}_\text{post}\left[ p_f(d_i\,|\,\bftheta)\right] \right) ,
\end{equation}
where we expressed integral over $p_f(d_i\,|\,\bftheta)$ in parameter
space as the expectation value $\mathbb{E}_\text{post}[\cdot]$ with
respect to the posterior distribution $p_f(\bftheta\,|\,\bfd)$ of the
parameters. We can proceed similar to the classical approach and
rescale with $-2N$ to obtain the Bayesian Predictive Information
Criterium\footnote{Albeit using a different approach for the
  derivation, this BPIC is similar to leave-one-out cross-validation
  \cite{vehtari:survey}.}
\begin{equation}
\label{eq:BPIC}
\text{BPIC}(f) = -2N\, \widehat{\kappa}(f)
\end{equation}
The model with the smaller value of the BPIC is favoured.
As discussed in section\,\ref{sect:bayes} the value of the BPIC 
depends on the chosen prior.
The expectation
$\mathbb{E}_\text{post}\left[p_f(d_i\,|\,\bftheta)\right]$ can be
evaluated with Markov Chain Monte Carlo (MCMC) methods. We start with
one chain simulating draws from $p_f(\bftheta\,|\,\bfd)$. Only this
chain is needed to calculate an estimate of
$\mathbb{E}_\text{post}\left[ p_f(d_i\,|\,\bftheta)\right]$ for each
of the data points $d_i$.
Contrary to section\,\ref{sec:information-classic} we do not use a
point estimate in the calculation of $\widehat{\kappa}(f)$.  Therefore
we think that a bias is not important in the calculation of
the BPIC$(f)$. Similar biases in the closely related leave-one-out
cross-validation are also considered negligible \cite{vehtari:survey}.

\subsection{Other methods}

A few other methods for model selection are in use.
To compensate the shortcomings of ordinary $p$--values
\cite{wasserstein:asa}, posterior \cite{meng:posterior,gelman:two} or
calibrated $p$--values \cite{sellke:calibration} have been suggested.
Closely related to the Bayes factor is the relative belief ratio which
measures the belief gained over the prior after an observation (see
\cite{baskurt:hypothesis}). Seehars et~al.\,\cite{seehars:information}
use the KL-divergence to quantify the information gained from new data
sets and define the ``surprise''.
The Deviance Information Criterium (DIC) was constructed by
Spiegelhalter et al.\,\cite{spiegelhalter:bayesian} as a revised
version of the AIC (see \cite{gelman:understanding} and
\cite{kunz:measuring}).  Although the DIC is popular, there is some
criticism (see \cite{spiegelhalter:12years} for a summary).
The derivation of the AIC \cite{akaike:information}, as sketched in
section\,\ref{sec:information-classic}, assumes that the parametric
models are regular\footnote{A statistical parametric model is regular
  if its Fisher matrix is positive definite.}. For typical
applications in cosmology, as given in section~\ref{sec:application},
this is the case.  However models defined by multilayered neural
networks are generically singular. For singular models Watanabe
derived the Widely Applicable Information Criterium (WAIC,
\cite{watanabe:asymptotic}).
With cross-validation we split the data set. A training sample is used
to determine the optimal parameters of the model and the remaining
part (the validation sample) is used for estimating the discrepancy
between the optimised model and the data.  Then one selects the model
with the smallest discrepancy (see \cite{arlot:survey} for a survey).
Gelman et al.\,\cite{gelman:understanding} compare the AIC, DIC, WAIC
and cross-validation in a variety of situations.
If one is interested in the estimates and the uncertainties of common
parameters in nested or overlapping models, Bayesian model averaging
could be a solution \cite{hoeting:bayesian,wassermann:bayesian}.
In the introduction we mention Ockham's razor and the principle of
parsimony.  This can be formalised by assigning the algorithmic
complexity as a unique measure to the model describing the data
\cite{kolmogorov:problems}.  Typically one is not able to calculate
the algorithmic complexity, but one can estimate the so called minimum
description length (MDL, \cite{rissanen:mdl}).  In its asymptotic form
the MDL is similar to the AIC and BIC with yet another Ockham's razor
term.
The approach from complexity theory and from information theory seem
to be closely related, but this is an open issue (see also
\cite{hansen:model}).

\section{An application -- the expansion history of the Universe}
\label{sec:application}

We illustrate these approaches to model selection by a classical
example from cosmology: the accelerating expansion of the Universe as
determined from redshift and luminosity measurements of supernovae
\cite{nobel:2011}. The question we address is whether this data allows
for a more detailed look at the expansion history of the universe and
specifically if we can decide between the $\Lambda$CDM and the $w$CDM
model.

A supernova~type\,Ia (SN\,Ia) is a stellar explosion with a well
defined luminosity \cite{hillebrandt:typeIa}.  In astronomy the
absolute luminosity is typically specified in logarithmic units, the
absolute magnitude $M_B$ in a given frequency range, here the
B-band. The observed flux is measured by the apparent magnitude $m_B$
(again in logarithmic units).  The distance modulus is defined as
$\mu:=m_B-M_B$.  The redshift $z$ of the supernova or of the hosting
galaxy is measured spectroscopically.
In a homogeneous and isotropic universe model the distance modulus --
redshift relation can be calculated.  Depending on the matter content
of the Universe we get
\begin{equation}
\mu(z, \bftheta) = 5 \log_{10} d_L(z, \bftheta) + 25 ,
\end{equation}
with luminosity distance $d_L$ in Mpc and the redshift $z$ of the
supernova.  The model dependence enters through the luminosity
distance $d_L(z, \bftheta)$ with the parameters $\bftheta$.
We do not pursue an exhaustive investigation of the currently
fashionable cosmological models and therefore fix some of the
otherwise free parameters.
Consider Nicola\,et\,al.\,\cite{nicola:consisteny} and Raveri\,\&\,Hu
\cite{raveri:concordance} for a comparison of more models using
comprehensive data sets.
We assume a spatially flat Universe ($\Omega_k=0$) and choose $H_0 =
70\text{km} \text{s}^{-1} \text{Mpc}^{-1}$ compatible with the data
from the Union\,2.1 sample \cite{suzuki:hubble}, but slightly larger
than the Planck value \cite{planck:planck15}.  Also if we consider
only supernovae, the value of the Hubble parameter is completely
degenerate with the absolute magnitude.
The overall scale is given by the Hubble distance
$d_H=\frac{c}{H_o}=4.28$\,Gpc, with $c$ the speed of light.
We consider two cosmological models:
\begin{compactenum}[1)]
\item The $\Lambda$CDM model with one parameter, the dimensionless
  density parameter $\Omega_m$.  Since we assume a flat background we
  have $0\le\Omega_m\le1$ and $\Omega_\Lambda=1-\Omega_m$ for the
  cosmological constant term.  The luminosity distance is given by
  (see e.g.\,\cite{hogg:distance})
  \begin{equation}
    d_L(z,\Omega_m) = d_H\,(1+z)\,
    \int_o^z \frac{\rmd z'}{\sqrt{\Omega_m(1+z')^3 + \Omega_\Lambda}} .
  \end{equation}
\item The flat $w$CDM model, with a constant equation of state
  $p=w\varrho$ for the dark energy component, has two free parameters
  $(\Omega_m, w)$. The density parameter obeys $0\le\Omega_m\le1$ and
  the cosmological term $\Omega_\Lambda=1-\Omega_m$ at present.  The
  time dependence of the cosmological term is parametrised using $w$
  and the luminosity distance is then
  \begin{equation}	
    d_L(z,\Omega_m,w) = d_H\,(1+z)\, \int_o^z
    \frac{\rmd z'}{\sqrt{\Omega_m(1+z')^3 + \Omega_\Lambda (1+z')^{3(1+w)} }} \ .
  \end{equation} 
\end{compactenum}

The observational data $d_i=(z_i,\mu_i)$ are the redshift $z_i$ and
the distance moduli $\mu_i$ of SN\,Ia.  In the Union\,2.1 sample we
have $N=580$ such observations from SN\,Ia together with an estimate
$\sigma_{\mu,i}$ of the uncertainty of each distance modulus
\cite{amanullah:spectra,suzuki:hubble}.
Assuming a cosmological model we calculate the distance modulus
$\mu_i$ given the redshift $z_i$ depending on the cosmological
parameters of the model.
The likelihood is the starting point for all the approaches to model
selection we discussed. Similar to eq.\,(\ref{eq:gauss-likelihood}) we
assume a Gaussian likelihood. For the flat $\Lambda$CDM model we have
\begin{equation}
  p_\Lambda(\bfd\,|\,\Omega_m) 
  = \frac{1}{\sqrt{(2 \pi )^N \prod_{i=1}^N \sigma_{\mu,i}^2 }} \ 
  \exp\left(- \frac{1}{2} \sum_{i=1}^N
  \frac{(\mu_i - \mu(z_i,\Omega_m))^2}{\sigma_{\mu,i}^2} \right)
\end{equation}
with $\bfd=(z_i,\mu_i)_{i=1}^N$ and the distance modulus
$\mu(z_i,\Omega_m)$ calculated from the model.  
This likelihood with a diagonal covariance matrix and model
independent variances is a simplification.  Our goal here is to
provide an illustrative example for the different approaches to model
selection. In appendix\,\ref{sec:errors} we will discuss the
statistical errors and systematic biases in more detail.
For the marginalised likelihood evaluated at $d_i$ we have from
eq.\,(\ref{eq:marginallikelihood})
\begin{equation}
  p_{\Lambda}(d_i\,|\,\Omega_m) 
  =  \frac{1}{\sqrt{2 \pi \sigma_{\mu,i}^2 }} \ 
  \exp\left(- \frac{1}{2}
  \frac{(\mu_i - \mu(z_i,\Omega_m))^2}{\sigma_{\mu,i}^2} \right) .
\end{equation}
The likelihood $p_w(\bfd\,|\,\Omega_m,w)$ and marginalised likelihood
$p_{w}(d_i\,|\,\Omega_m,w) $ of the $w$CDM model are defined
analogously.  In the Bayesian analysis we need to specify the
priors. We assume a uniform distribution on $[0,1]$ for $\Omega_m$ and
a uniform distribution on $[-2,0]$ for $w$.
\begin{figure}
  \centering
  \includegraphics[width=0.8\linewidth]{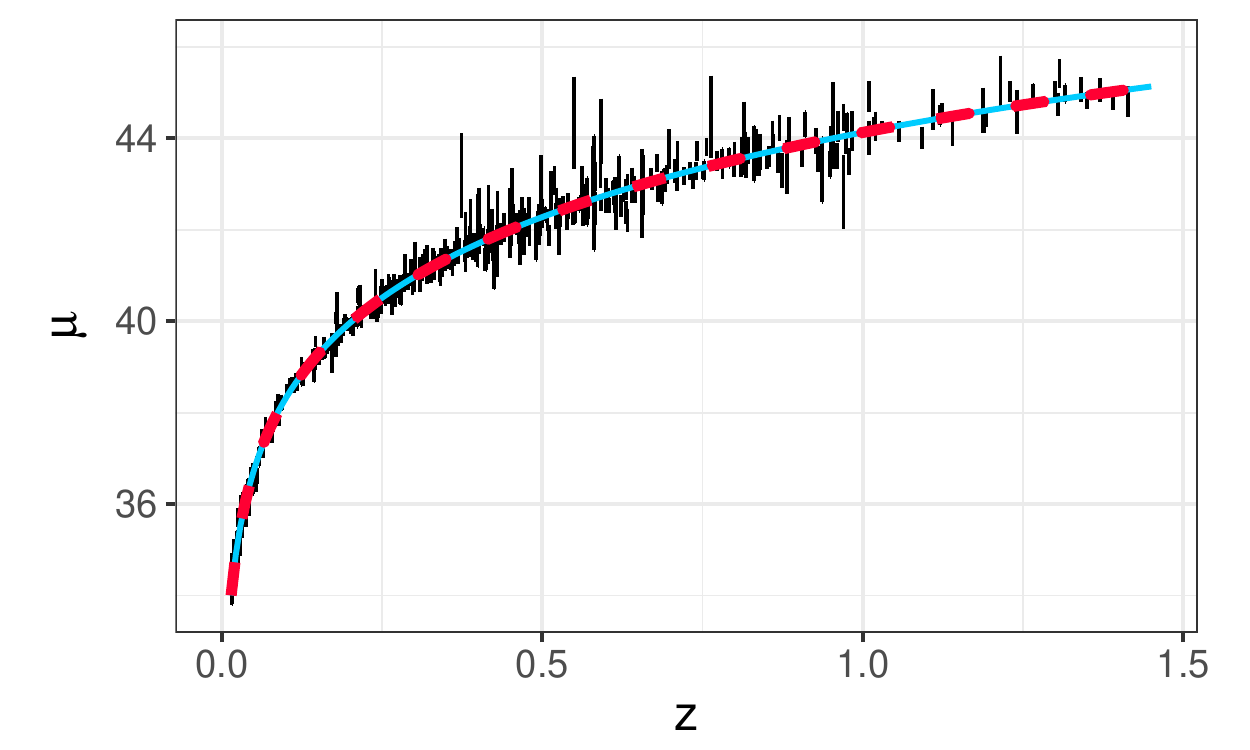}
  \caption{The redshift $z$ plotted versus distance modulus $\mu$ for
    the SN\,1a data together with the fits for the $\Lambda$CDM model
    (solid blue line) and the $w$CDM model (dashed red line).}
  \label{fig:datafits}
\end{figure}
Before we turn to model selection, we estimate the parameters.  For
the flat $\Lambda$CDM model we obtain $\Omega_m=0.278\pm0.007$. This
estimate is virtually identical between the least square fit, the
maximum likelihood and the MAP estimate. The error shown is the
standard deviation of the posterior distribution. Similarly, we obtain
$\Omega_m=0.279\pm0.06$ and $w=-1.0\pm0.13$ in the $w$CDM model.
Remember, we fixed $H_0$ and only consider spatially flat cosmological
models.
In figure\,\ref{fig:datafits} we show the data points together with
the prediction of the two models, using the best fit values of the
parameters respectively. In this plot the curves from the two models
are lying indistinguishably on top of each other.

Now we apply all the methods for model selection described previously.
In appendix~\ref{sec:implement} we give details on the numerical
procedures used. First we summarise the results, later on we will put
them in perspective by investigating the fluctuations of the results.
\begin{compactdesc}
\item{\bf Goodness of fit:} In the $\Lambda$CDM model we have a
  $\chi^2_\text{red}(\Lambda)=0.971$ resulting in a $p$--value of
  $0.68$ and in the $w$CDM model a $\chi^2_\text{red}(w)=0.973$ with a
  $p$--value of $0.67$. Neither the $\Lambda$CDM nor the $w$CDM model
  can be rejected.
\item{\bf Likelihood ratio test:} From the maxima of the likelihoods
  of both models we compute the likelihood ratio $L$ and then
  $\lambda=-2\log L= 0.000948$ . This results in a $p$--value of
  $1-G_1(\lambda) = 0.975$. Clearly we cannot reject the $\Lambda$CDM
  in favour of the $w$CDM model.
\item{\bf Bayesian approach:} For the $\Lambda$CDM model we obtain a
  $\text{BIC}_\Lambda=-231.1$ and for the $w$CDM model a
  $\text{BIC}_w=-224.8$. Hence we should prefer the $\Lambda$CDM
  model.  The evidence ratio is $B_{\Lambda w} =
  \frac{p_\Lambda(\bfd)}{p_w(\bfd)}= 5.45>1$, and again we should
  prefer the $\Lambda$CDM model.
\item{\bf Classical information theoretic approach:} For the
  $\Lambda$CDM model we have an $\text{AIC}_\Lambda=-235.5$ and for
  the $w$CDM model an $\text{AIC}_w=-233.5$. Hence we should prefer
  the $\Lambda$CDM model.
  For the $\Lambda$CDM model we get 
  $\text{EIC}_\Lambda=-239.3$ 
  and for the $w$CDM model 
  $\text{EIC}_w=-241.0$.  
  This suggests that we should prefer the $w$CDM model.
\item{\bf Bayesian information theoretic approach:} We obtain a
  $\text{BPIC}_\Lambda=-237.5$ 
  for the $\Lambda$CDM model and a
  $\text{BPIC}_w=-237.3$ 
  for the $w$CDM model. Therefore the $\Lambda$CDM model is 
  preferred over the $w$CDM model.
\end{compactdesc}

\subsection{Stability of the model selection}
Neither using the least-square results nor with the likelihood ratio
test we arrive at a definite conclusion. Both models fit the data, and
we also cannot rule out $\Lambda$CDM or the $w$CDM model.
In the Bayesian approach often Jeffereys' \cite{jeffreys:theory} scale
is employed to express the numerical value of the evidence ratio
$B_{\Lambda w}$ in words\footnote{Jeffreys' scale for the evidence
  ratio $B$ translated to our conventions reads (see appendix\,B
  in\,\cite{jeffreys:theory}): $B<1$: negative evidence; $1\le
  B<\sqrt{10}$: barely worth mentioning; $\sqrt{10}\le B<10$:
  substantial; $10\le B<10^{3/2}$: strong; $10^{3/2}\le B<100$: very
  strong, $100\le B$: decisive evidence. \label{foot:jeffreys}}.
Hence with $B_{\Lambda w}=5.4$ we have ``substantial evidence'' to
support the $\Lambda$CDM over the $w$CDM.  However such a
``universal'' scale is disputed (see e.g.\ \cite{efron:scale} or
\cite{nesseris:jeffreys}).
Similarly, the mere comparison of numbers, like we did with the AIC,
BIC, EIC and BPIC, is not satisfying.  A scale is missing.

\begin{table}
  \caption{A summary of the results from the Union\,2.1 data set
    together with the dispersion estimated from the $\Lambda$CDM mock
    samples as described in the text.  \label{table:dispersion}}
\begin{center}
\begin{tabular}{|l|r|r|r|}
  \hline
  & results from the & difference            & ``$\Delta$'' from \\
  & Union 2.1 sample     & ``$|\Lambda-w|$'' & the mocks\\
  \hline
  $p_\Lambda$ &  0.68  & \multirow{2}{*}{0.01} &  \multirow{2}{*}{0.458 
                                                 \tablefootnote{One can show 
                                                 that the $p$--values obtained 
                                                 from these mock-samples are 
                                                 uniformly distributed on $[0,1]$ 
                                                 and therefore the ``$\Delta$'' 
                                                 is not really informative.\label{foot:pval}}} \\
  $p_w $           &  0.67  &  &  \\
  \hline
  $p$ likelihood ratio &  0.975  &  --- & 0.565 \footref{foot:pval}\\
  \hline
  $B_{\Lambda w}$ &  5.45 &  --- & 3.41 \\
  \hline
  $\text{BIC}_\Lambda$ &  -231.1  & \multirow{2}{*}{6.3} &  \multirow{2}{*}{42.2} \\
  $\text{BIC}_w$            &  -224.8  & & \\
  \hline
  $\text{AIC}_\Lambda$ &  -235.5  & \multirow{2}{*}{2.0} &  \multirow{2}{*}{42.2} \\
  $\text{AIC}_w$            & -233.5 & & \\
  \hline
  $\text{EIC}_\Lambda$ &  -239.3 & \multirow{2}{*}{$1.7$} &  \multirow{2}{*}{$39.4$} \\
  $\text{EIC}_w$            & -241.0  & & \\
  \hline
  $\text{BPIC}_\Lambda$ &  -237.5  & \multirow{2}{*}{$0.2$} &  \multirow{2}{*}{$42.9$} \\
  $\text{BPIC}_w$            &  -237.3 & & \\
  \hline
%
\end{tabular}
\end{center}
\end{table}

We do not want to propose a universal scale, which probably does not
exist, but we suggest a model dependent approach to investigate the
stability of our model selection.
As a concrete example, consider the Bayesian information theoretic
approach and the values of $\text{BPIC}_\Lambda=-237.5$ and
$\text{BPIC}_w=-237.3$ for the $\Lambda$CDM and the $w$CDM model,
respectively.
To see whether this difference is important we repeatedly generate
artificial data sets and calculate the $\text{BPIC}_\Lambda$ for each
of these data sets. This allows us to estimate the dispersion
$\Delta_{\text{BPIC}_\Lambda}$. Clearly the fluctuations depend on how
we generate our artificial data set. We start with the Union\,2.1
sample \cite{suzuki:hubble} and keep the redshift $z_i$ and the
uncertainty $\sigma_{\mu,i}$ fixed and generate generate randomised
distance moduli $\widetilde{\mu_i}$ for each of the supernovae, The
$\widetilde{\mu_i}$ fluctuate around the prediction of the
$\Lambda$CDM model according to
\begin{equation}
\widetilde{\mu_i} = 5 \log d_L(z_i,\Omega_m=0.278)+25 + s_i ,
\end{equation}
where $s_i$ is a random number, normally distributed with zero mean
and standard deviation $\sigma_{\mu,i}$.  This gives us an artificial
data set $(z_i, \widetilde{\mu_i}, \sigma_{\mu,i})_{i=1}^{580}$.
For one hundred of these artificial data sets we calculate the BPIC
and estimate the dispersion of the BPIC using the
mid-spread\footnote{The mid-spread, or inter quartile range, is
  defined as the difference between 75th and 25th percentiles. It is a
  robust estimator of dispersion. For a Gaussian distribution the
  mid-spread is approximately $1.34$ times the standard deviation.}
$\Delta_{\text{BPIC}_\Lambda}=42.9$.
This dispersion estimate $\Delta_{\text{BPIC}_\Lambda}$ is two
orders of magnitude larger than the difference between
$\text{BPIC}_\Lambda$ and $\text{BPIC}_w$. Hence, using the BPIC  we
can not select one of the models.
This is not a full evaluation of the fluctuations present in the
model, but it helps us to assess the relevance of our results in a
model dependent way (see also appendix\ref{sec:errors}).  
The results from the Union\,2.1 sample and the dispersion estimates
for the $p$--values, the $B_{\Lambda w}$, BIC, AIC, EIC, BPIC are
summarised in table\,\ref{table:dispersion}. The dispersions
``$\Delta$'' are always significantly larger than the observed
differences between the $\Lambda$CDM and $w$CDM models. Fixing levels
or using universal scales for the various criteria can hence be
misleading (see also
\cite{raveri:concordance,colquhoun:reproducibility}).

\begin{figure}
  \centering
  \includegraphics[width=0.24\linewidth]{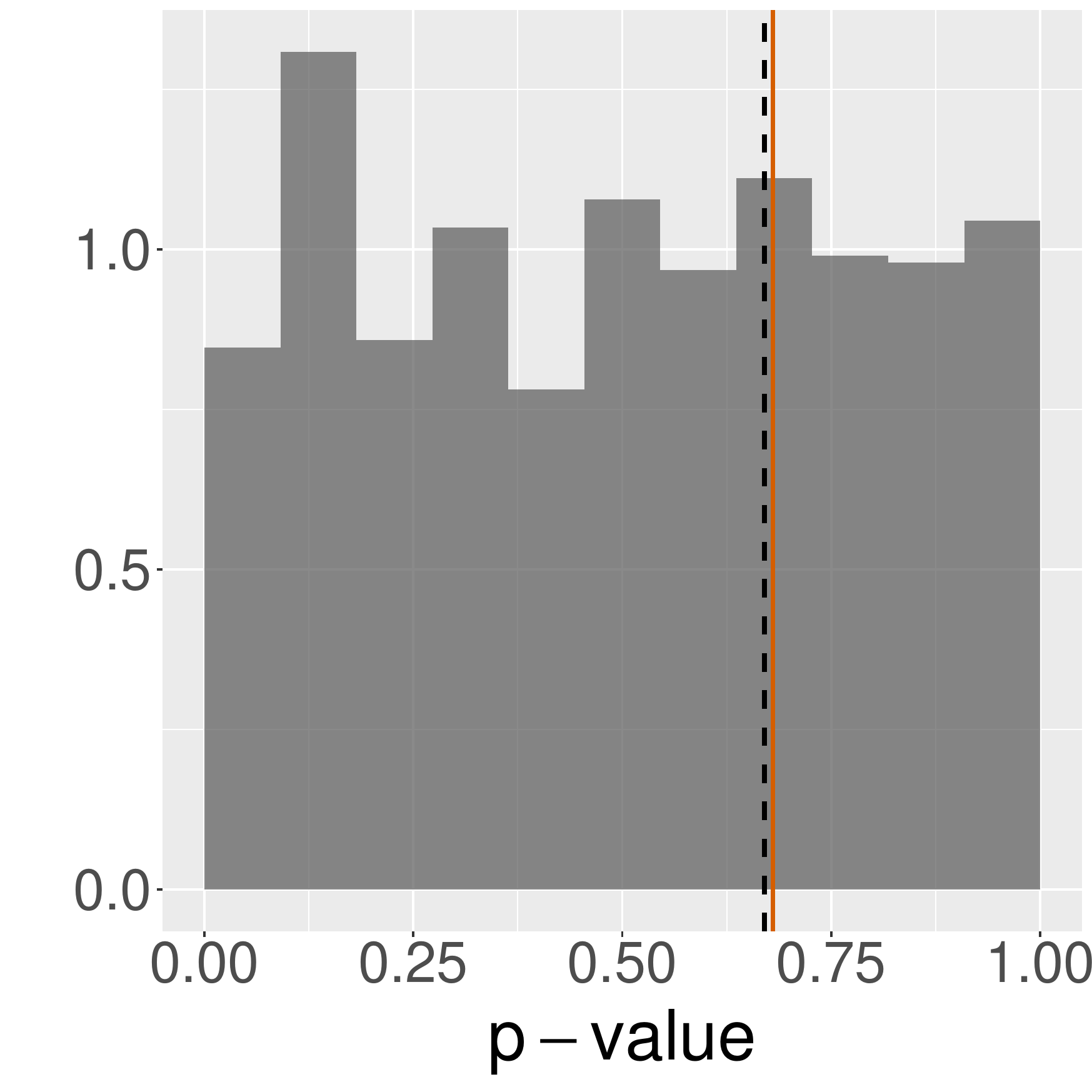}\hfill
  \includegraphics[width=0.24\linewidth]{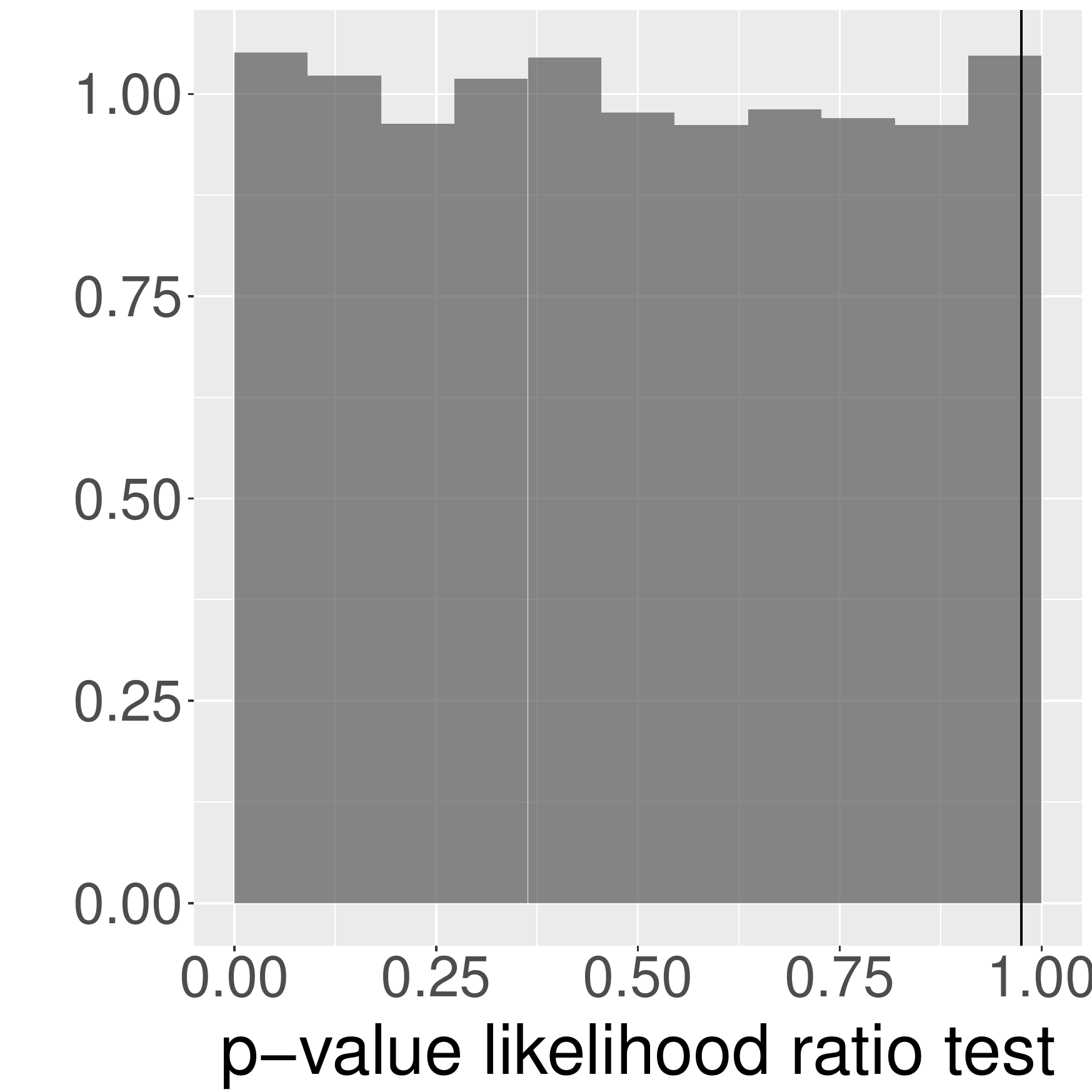}\hfill
  \includegraphics[width=0.24\linewidth]{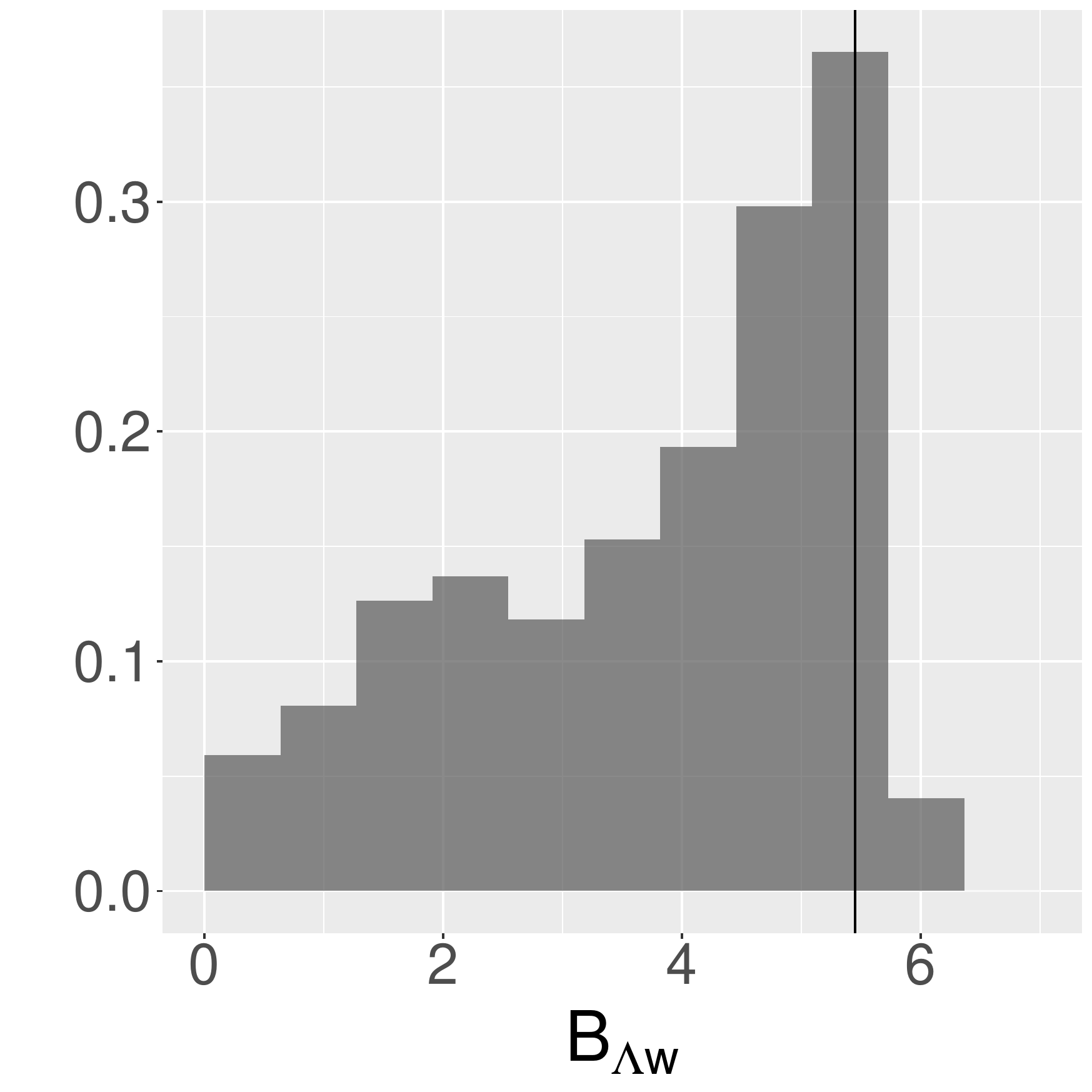}\hfill
  \includegraphics[width=0.24\linewidth]{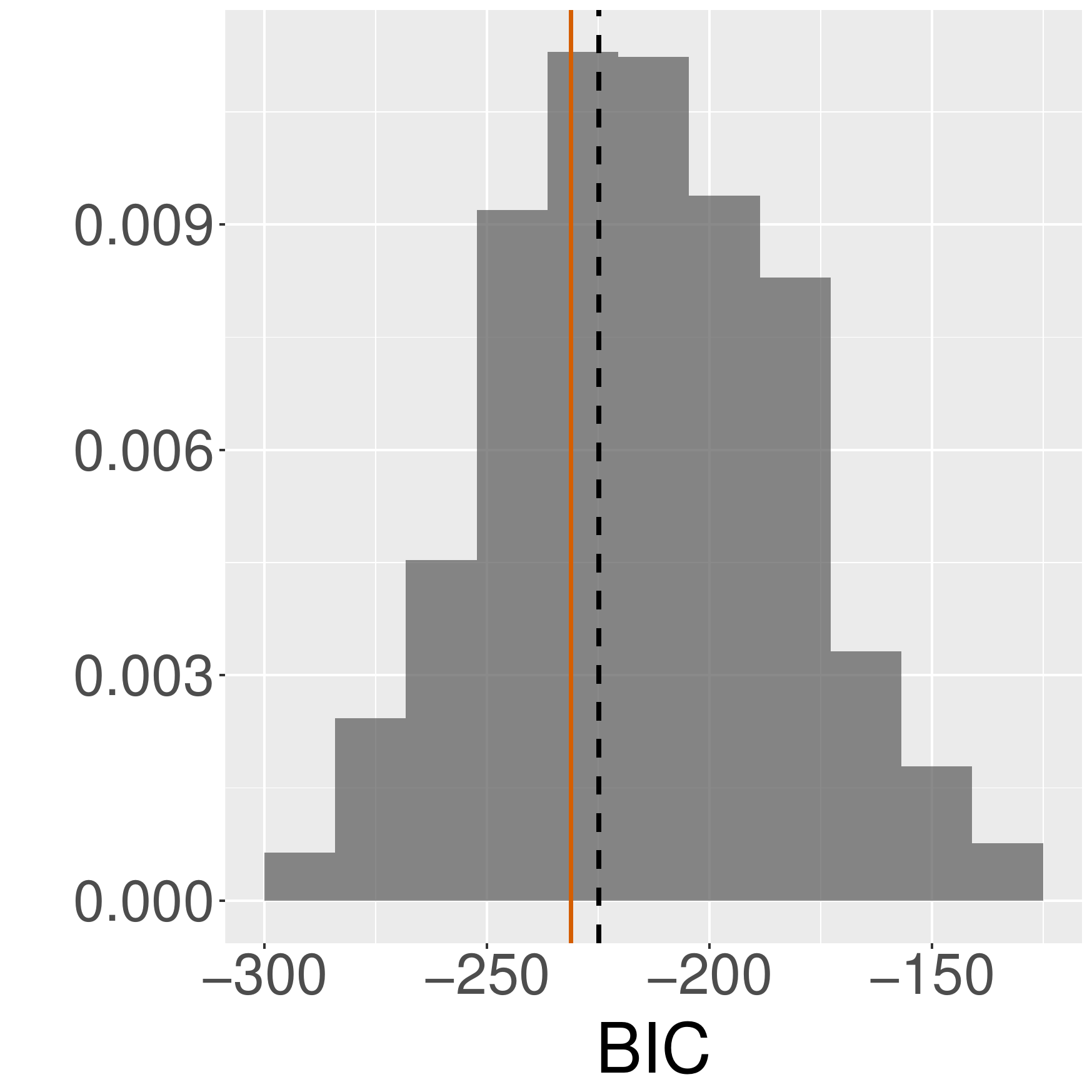}\\[1em]
  \includegraphics[width=0.24\linewidth]{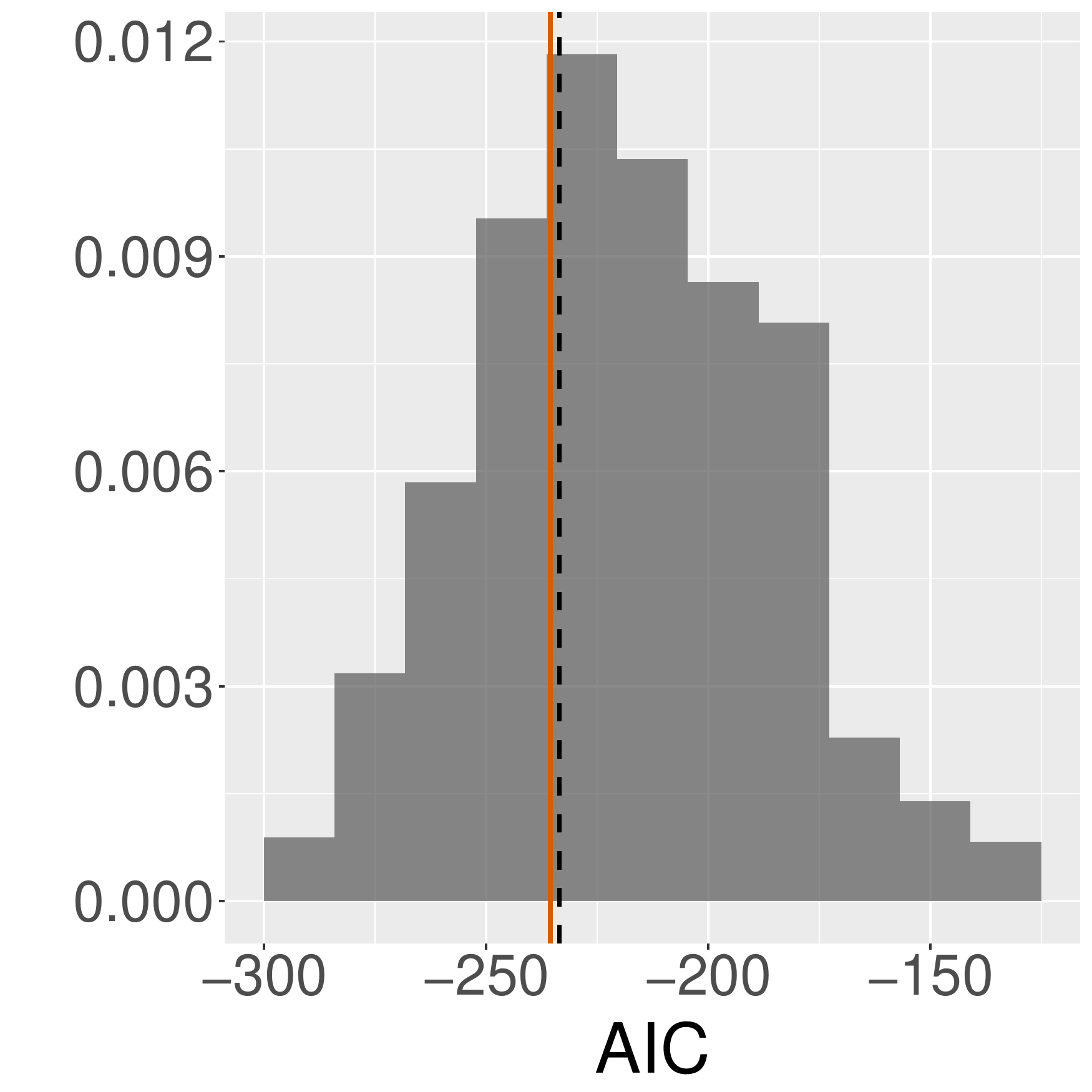}\quad
  \includegraphics[width=0.24\linewidth]{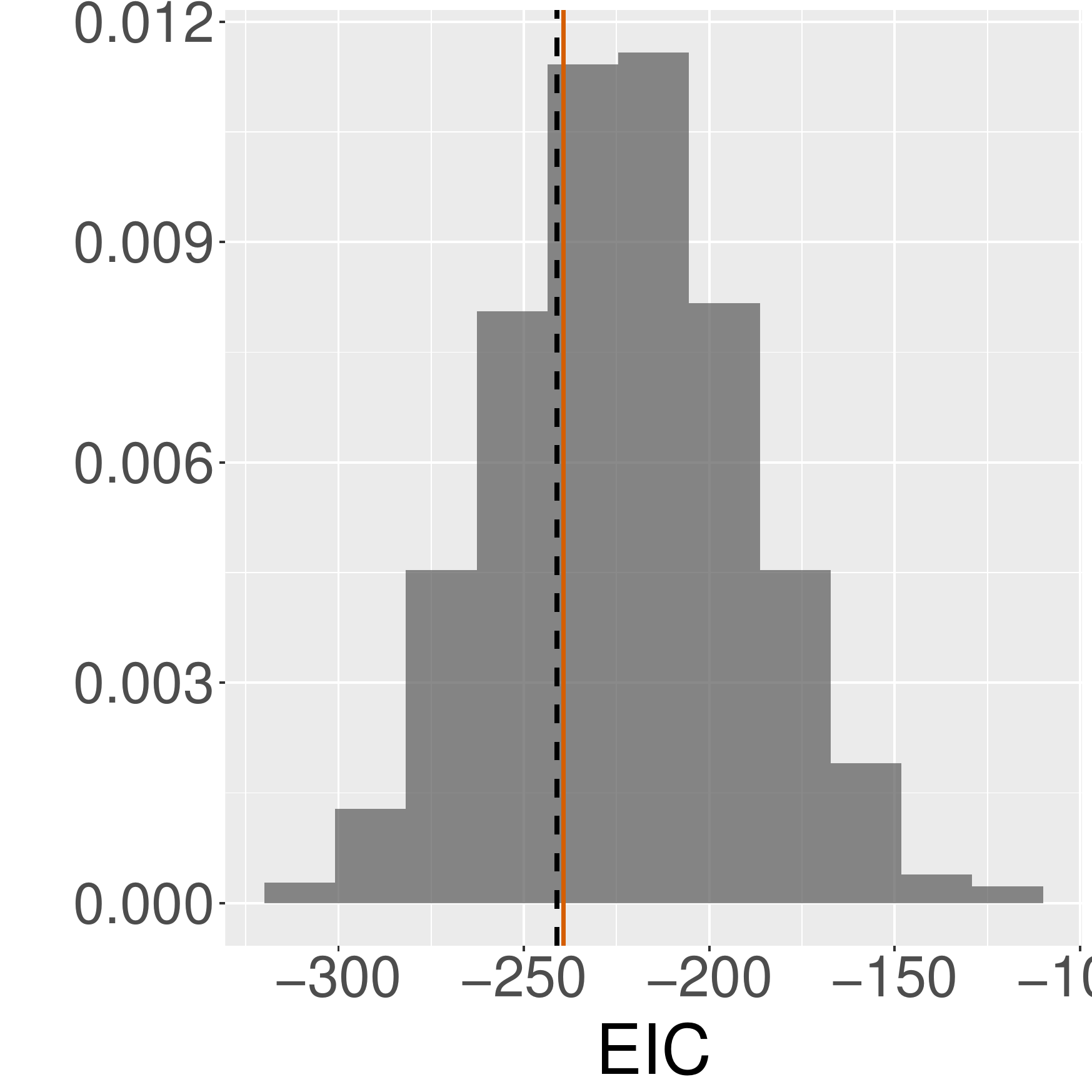}\quad
  \includegraphics[width=0.24\linewidth]{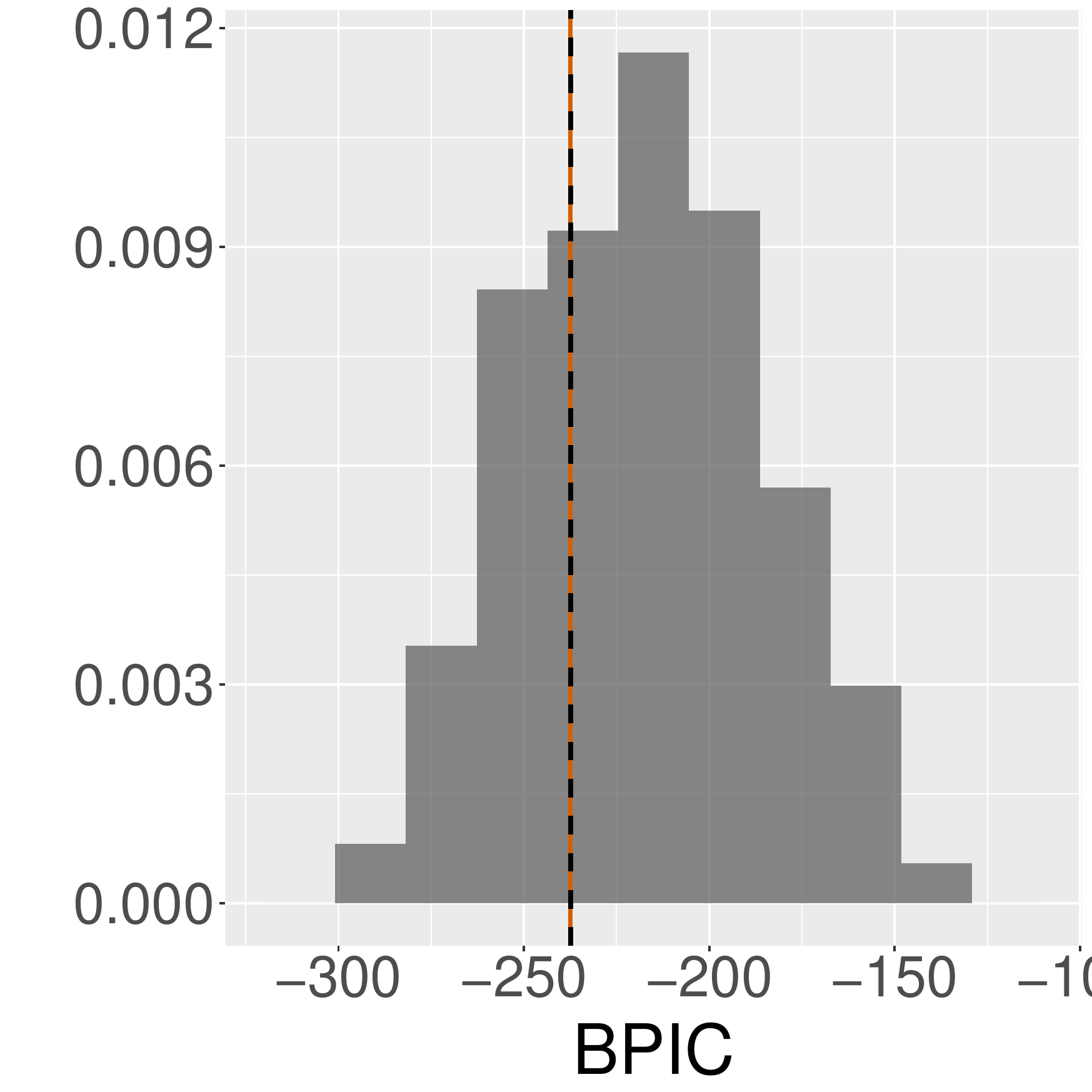}
  \caption{The normalised frequency distribution of the relevant
    quantities for model selection obtained from (at least)
    100\,$\Lambda$CDM mock samples. The vertical lines mark the values
    obtained from the Union\,2.1 sample (compare
    table\,\ref{table:dispersion}) for the $\Lambda$CDM (red) and
    $w$CDM model (dashed black). The $p$-value from the likelihood
    ratio test and the Bayes factor are already comparative quantities
    and only one black vertical line is shown. }
    \label{fig:histogramms}
\end{figure}

We may considers the data as a realisation of a random process.  Then
it is quite natural to quantify the dispersions in this model
dependent way. All the key figures are random variables depending on
the model and the data set (considered as a random realisation). As a
showcase we give the empirical distributions of all the relevant
quantities for model selection within this $\Lambda$CDM mock scenario
in figure\,\ref{fig:histogramms}.  But one should be aware, that in
classical statistics such a mock scenario is
unnatural\footref{foot:pval}.  The $p$--value is considered a fixed
number, only depending on the data set under investigation and the
likelihood.

As mentioned in section\,\ref{sect:bayes} we study the dependence on
the priors.  We calculate the Bayes factor $B_{\Lambda w}$ and the BPIC 
for a series of priors. Still we restrict the
parameters to the ranges $\Omega_m\in[0,1]$ and $w\in[-2,0]$, but in
addition to the flat distribution we use Jeffreys' prior (a suitably
rescaled Beta(1/2,1/2) distribution) and a series of truncated and
renormalised Gaussian distributions. For the truncated Gaussian
distributions we vary the width from almost flat on the intervals to
strongly peaked and we use two different mean values, one is centred
on the ``correct'' value (the MAP estimate).
For all these priors the posterior distributions are very similar and
the MAP estimates agree within the fluctuations.
The $\text{BPIC}_{\Lambda}$ from the Bayesian information
theoretic approach ranges from -237.4450 to  -237.4485 for these priors.
Only in the extreme cases, where we have negligible overlap between
the prior and the posterior distribution, we get a value outside this
range.
Hence, if the prior is sufficiently broad and shows some overlap with
the posterior distribution we get consistent results for the BPIC
irrespective of the prior.  A similar behaviour is observed for the
Bayes factor $B_{\Lambda w}$.

Our conclusion comes as little surprise (see
e.g.\,\cite{trotta:probing}): using the Union\,2.1 data set we cannot
decide whether the $\Lambda$CDM or the $w$CDM model should be
preferred.  Also keep in mind that we use a simplified ansatz for the
likelihood. See appendix\,\ref{sec:errors} for further notes.

\section{Discussion of the methods}
\label{sec:discussion}

In physics the construction of models is guided by basic principles
(conservation laws, symmetries, etc.). Adding another term, as
illustrated in Fig.\,\ref{fig:numquam} in the introduction, is often
not acceptable because one would violate these principles.
For statistical applications in engineering or the social sciences
this is often not a major concern. Model selection is used as a
criterium to decide whether one should introduce new parameters and
new dependencies.
Ockham's razor suggests that one should go for the simpler model.
However simplicity needs to be quantified (see
Sober\,\cite{sober:simplicity}).  The dimension of the parameter space
immediately comes to mind as such a measure of simplicity.  But the
dimension is only a rough and sometimes misleading measure of
parsimony (see e.g.\ \cite{sober:evidence}, and also compare
Fig.\,\ref{fig:numquam}).
The goodness of fit, the likelihood ratio, the evidence ratio, or the
KL-divergence from the information theoretic approaches are
operationally well defined procedures for model selection. They allow
quantitative arguments beyond mere qualitative arguments.
In section\,\ref{sec:modelcompare} we describe these methods and in
section\,\ref{sec:application} we apply them to a problem from
cosmology. Typically one would not want to apply all of them. Neither
from the mathematical definitions nor from the data analysis a clear
recommendation emerges.  We will now present some philosophical
arguments and finally recommend the information theoretical approach.
First the methods, given in Sect.~\ref{sec:modelcompare}, are briefly
summarised before we critically discuss them:
\begin{compactitem}[-]
\item With the goodness of fit one ranks models according to 
  their ability to fit the data points.
\item With the likelihood ratio you compare the probabilities of your
  data given the best fitting models. Together with a predefined
  significance level the likelihood ratio allows you to discard a
  given model (your null~hypothesis) in favour of the alternative
  model.
\item In a Bayesian model comparison you use the evidence ratio to
  compare the joint probabilities of the models and the data. This
  depends on the likelihood and the prior.
\item In the classical information theoretic approach you measure how
  good the best fitting models are at predicting new data.
\item In the Bayesian information theoretic approach you measure how
  good the posterior predictive distributions of the models are at
  predicting new data.
\end{compactitem}

The ``goodness of fit'' based on the $\chi^2_{f,\text{red}}$ is
sometimes used for model selection.  The major shortcoming is that the
$\chi^2_{f,\text{red}}$ does not factor in any contributions from the
false negative rate (compare appendix\,\ref{sect:stat}).  If we
specify a second model and assume a Gaussian error model as well as
independent sampling, the difference $\chi^2_{f}-\chi^2_{g}$ is
related to the likelihood ratio as used in the likelihood ratio test.

Although the likelihood ratio test and the Bayesian model selection
derive from quite different approaches towards statistical analysis,
they both assume that the true model is among the considered models
(see also \cite{vehtari:survey}). Then you either discard the false
models via tests, or you determine the most probable model.
The information theoretic approach is different. There one accepts
that a model is an approximation and one tries to identify the model
which is closest to the true empirical distribution. This approach
allows us to predict new data in the best possible way.

Similarly Wit et al.\,\cite{wit:models} discuss the following two
questions (see also \cite{vehtari:survey}): i) which modelling
procedure will, with sufficient data, identify the true model? or ii)
based on the data, which model lies closest to the true model?
They conclude indecisively: asking different questions leads to
different approaches for model selection.
However one is able to go beyond this neutral statement.  Consider the
following aphorism attributed to G.~Box\,\cite{box:science}: ``all
models are wrong''.  In physics one would not use the term
``wrong''. Physical models have their range of applicability. We know
that Newtonian gravity is failing on large scales and we assume that
general relativity is failing on very small scales. However both have
their range of applicability and we successfully compare their
predictions with measurements and observations.
Presumably all models in physics, at least the models which may be
confronted with data, are effective models (see for example the
discussion of effective field theories in \cite{hartmann:effective}).
Hence, methods of model selection, which try to identify the true
model are deceptive.  We know from the outset that our models are
``wrong''. This is a bit nitpicking, since we know about the range of
applicability of our models. Nevertheless it is advisable to respect
this situation from the beginning and use the information theoretic
approach. There we try to find the best approximate, not necessarily
the true model.
This becomes especially important in cosmology, where new observations
always add to the existing data. For example new observations of
galaxies are added to the already known galaxy catalogues.  The
Universe contains the galaxy distribution and probabilistic physical
models are used to describe it (see \cite{beisbart:probabilistic}).
Again, we seek the best approximating model.

Now consider another argument from the philosophy of science (see
also\cite{gelman:philosophy}).  Bayesian updating is sometimes
presented as the only relevant way of plausible reasoning in science
(Jaynes\cite{jaynes:brain}). This would favour methods based on the
Bayesian evidence and the evidence ratio for model selection.
However scientists devise new models and compare them to data. Either
the data supports the model or sometimes allows a rejection
(falsification). This cycle has been put forward by
Popper\cite{popper:logik} and refined by
Lakatosz\cite{lakatosz:methodology}. Actually this approach seems to
be too restrictive to describe the scientific growth of knowledge as
outlined by Feyerabend\cite{feyerabend:against} and
Kuhn\cite{kuhn:structure}. Laudan\cite{laudan:progress} argues that
the contextual problem solving effectiveness is the key ingredient for
a successful description of scientific progress (compare also
with\cite{kuhn:objectivity}).  In other words, one seeks the model
which offers the most effective way to describe new data.
This is the idea behind the information theoretic approach.

Up to now we argued for the information theoretic approach in general
but did not differentiate between the classical and the Bayesian
version.  As we already stated in the introduction we prefer the
Bayesian approach if the prior is well specified.  We do not want to
repeat the arguments exchanged in the discussion of the Bayesian
versus the classical approach to statistics. Perhaps the articles by
Cousins\,\cite{cousins:why} and Efron\,\cite{efron:why}, including the
comments directly following Efron's article, may serve as an
introduction to this discussion.  A pragmatic reconciliation is
suggested by Kass~\cite{kass:statistical} in his ``big picture''.

So far we presented methods for the comparison of models based on
their ability to fit or predict observational data. 
There are further criteria we can and should employ to assess physical
models.  Independent from the observational data, new models
(hopefully) make predictions and solve conceptual problems. They can
be judged by their effectiveness to solve such problems
\cite{laudan:progress}. This is not a quantitative endeavour, one has
to present arguments for and against models, often based on the
foundations of the models, or criticising the viability of the
approximations used. But in the end physical models have to stand the
comparison with data \cite{ellis:defending}.  Then the methods of
model selection we discuss come into play.

\section*{Acknowledgements}

Many thanks to all the discussants from the physical cosmology and the
OPINAS seminar at LMU and the Bayes forum in Munich which helped us to
improve and sharpen the argumentation.  MK wishes to thank Claus
Beisbart, Ulrich Schollwöck and Herbert Wagner for stimulating
discussions and helpful comments.


\begin{appendix}
\section{Some results from statistics}
\label{sect:stat}

\paragraph{Statistical tests}
The theory of hypothesis tests for statistical data analysis was
pioneered by K.\,Pearson\,\cite{pearson:criterion}.  His goal was to
compare the observed frequency distribution of random events to
probabilities from a model.  He could show that the test statistic he
developed, asymptotically follows a $\chi^2$--distribution.  This
approach was significantly extended by
Fisher\,\cite{fisher:statistical}.
We follow the practice in physics and name the mean-square calculated
in eq.\,(\ref{eq:chi2}) by $\chi^2_f$ (see e.g. \cite{press:recipes2}
or \cite[chap.~40]{patrignani:pdg}).  This $\chi^2_f$ is clearly
different from the test statistic used by
Pearson\,\cite{pearson:criterion}.
If we make the strong assumptions that the $N$ data points used in the
calculation of eq.\,(\ref{eq:chi2}) are independent and that the error
is Gaussian distributed, then $\chi^2_f$ is asymptotically following a
$\chi^2$--distribution with $N-1$ degrees of freedom (see also
\cite{andrae:dosanddonts}).

In this situation a statistical test proceeds in the following
way. Our null~hypothesis is that our model with the best parameter
$\bftheta^\starii$ from the least square fit is correct. Then the
$\chi^2_f$ is calculated using eq.\,(\ref{eq:chi2}). The $p$--value is
given by $p=1-G_{N-1}\left(\chi^2_f\right)$, where $G_\nu$ is the
cumulative distribution function of a $\chi^2$--distributed random
variable with $\nu$ degrees of freedom. The p--value is the
probability that the data may arise from the null~hypothesis (see
section\,\ref{sect:goodness} for more comments on the p--value).
\begin{figure}
  \centering
  \includegraphics[width=0.48\linewidth]{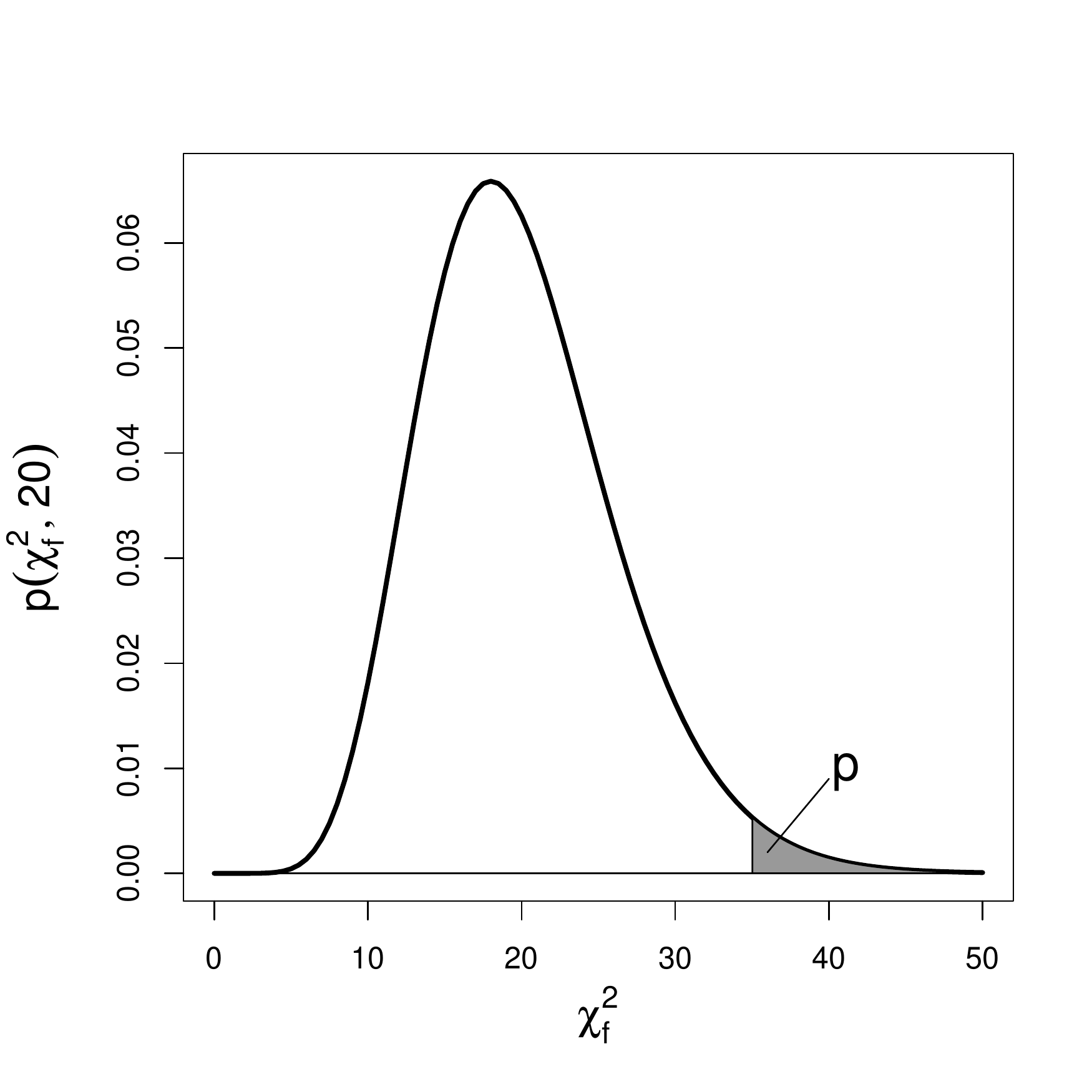}
  \hfill
  \includegraphics[width=0.48\linewidth]{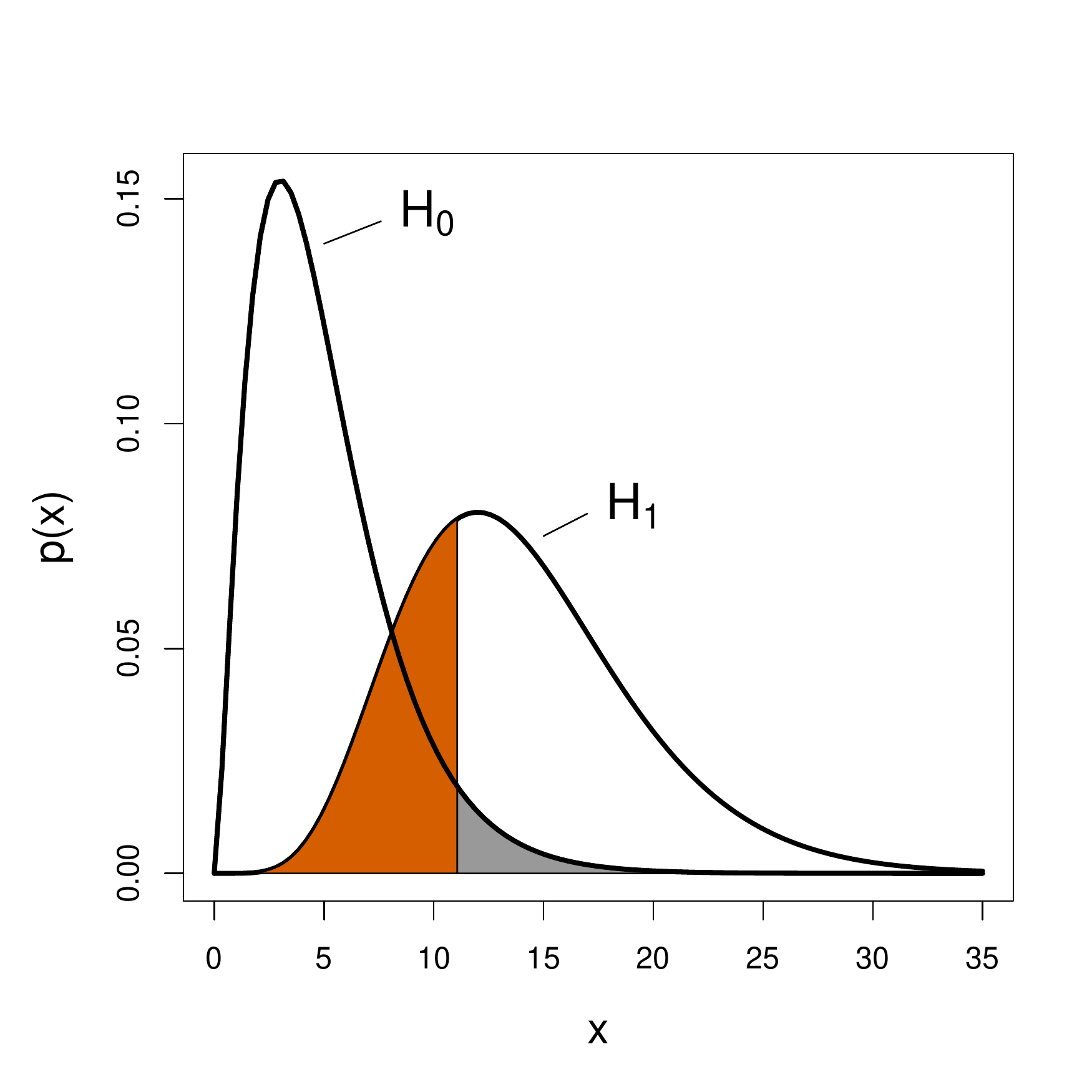}
  \caption{In the left plot the probability density
    $p\left(\chi^2_f,20\right)$ of a $\chi^2$--random variable with 20
    degrees of freedom is shown. The shaded area marks the 
    probability $p=0.02$ of obtaining a value of $\chi^2_f>35.0$.
    The right plot shows the probability densities of the
    null~hypothesis and the alternative hypothesis. In this situation
    we have a false positive rate $\alpha=0.05$ (grey) and a false
    negative rate $\beta=0.32$ (red).  }
    \label{fig:pvalue_secondkind}
\end{figure}
To round up the test we fix a so called significance level, typically
$\alpha=0.05$ (also called the ``false positive rate'' or type\,I
error).  If $p<\alpha$ we may conclude that the null~hypothesis (our
model) is rejected at the $\alpha=0.05$ significance level. The left
plot in figure\,\ref{fig:pvalue_secondkind} illustrates such a
situation.

For the calculation of the false negative rate, specifying the
null~hypothesis $H_0$ alone is not sufficient \cite{neyman:problem}.
We have to state an alternative model, the hypothesis $H_1$.  Now
assume that $H_1$ is true but $H_0$ has \emph{not} been rejected
(i.e.\ $H_0$ has been falsely accepted).  Given $H_1$ and the false
positive rate $\alpha$ we can calculate the false negative rate
$\beta$ (also called the type\,II error) as
illustrated in the right plot of Fig.~\ref{fig:pvalue_secondkind}.
In the goodness of fit approach one does not specify an alternative
model, the hypothesis $H_1$. Hence one is not able to quantify the
false negative rate. As we see in the right plot of
Fig.~\ref{fig:pvalue_secondkind} the false negative rate $\beta$ can
be quite large even for small $\alpha$.

\begin{figure}
  \centering
  \includegraphics[width=0.5\linewidth]{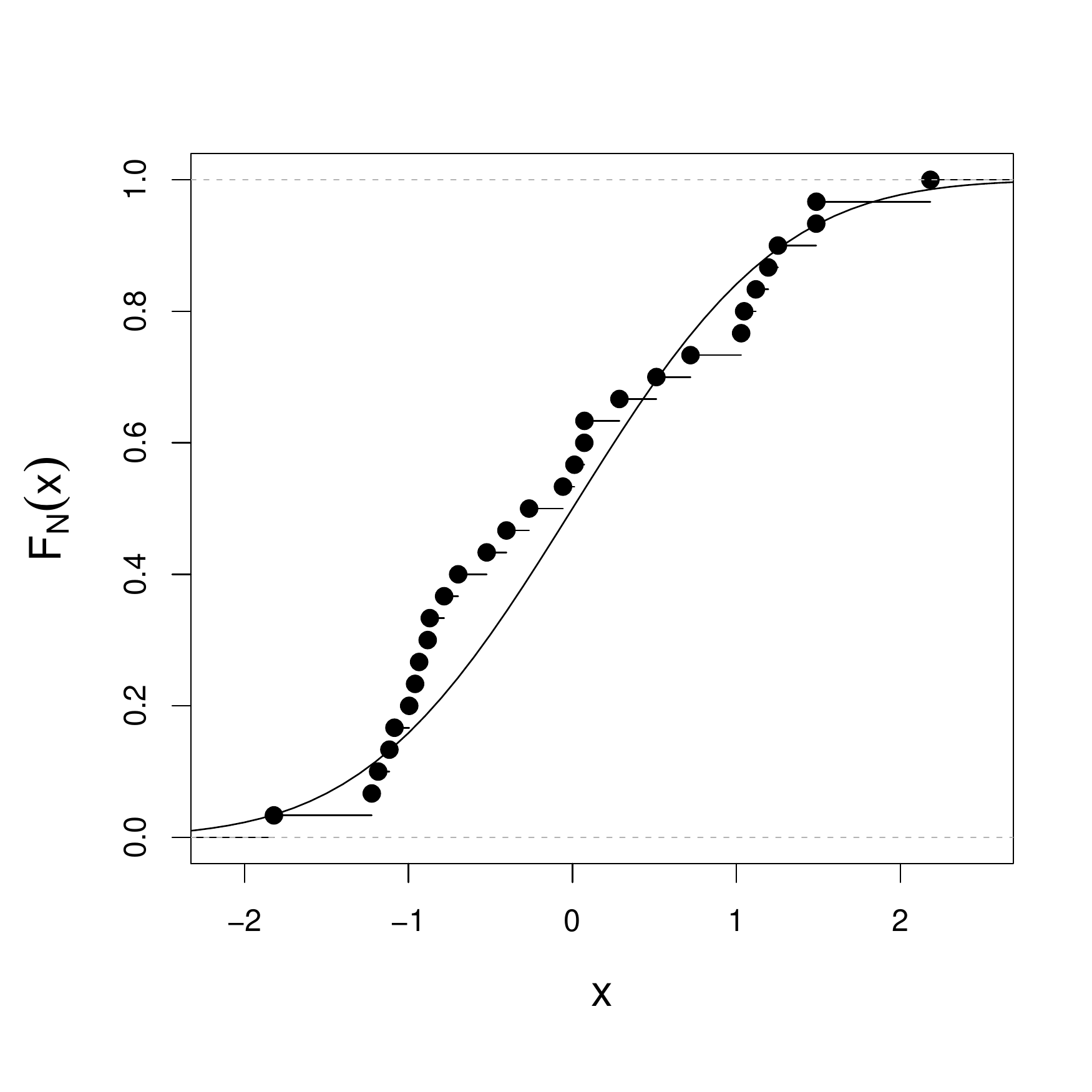}
  \caption{The cumulative distribution function of a standard Gaussian
    random variable together with the empirical distribution function
    for $N=30$ random realisations. }
    \label{fig:glivenko}
\end{figure}

\paragraph{Empirical distribution function:}
For simplicity consider a real valued random variable with probability
density $p(x)$ and cumulative distribution function
\begin{equation}
  F(x) = \int_{-\infty}^x p(x') \rd x' .
\end{equation}
Consider $N$ independent random realisations $(x_1,\ldots,x_N)$ of
this random variable. Then the empirical distribution function is
defined as
\begin{equation}
\label{eq:empiricaldist}
F_N(x) := \frac{1}{N} \sum_{i=1}^N I_{[x_i,\infty)}(x) .
\end{equation}
Here $I_A(x)$ is the indicator function of the set $A$ with $I_A(x)=1$
if $x\in A$ and zero for $x\not\in A$.  The theorem of
Glivenko--Cantelli states that $F_N(x)$ converges for
$N\rightarrow\infty$ towards $F(x)$ uniformly, this means
\mbox{$\| F_N-F\|_\infty\rightarrow0$} almost surely (see 
\cite{billingsley:probability}, and compare figure~\ref{fig:glivenko}).
For example this theorem guarantees the convergence of
the empirical median and quantiles.
The empirical distribution function is analogously defined in higher
dimensions.  The half--interval is replaced by a half open rectangle
stretching to infinity with the point $x_i$ marking the lower left
corner.

\paragraph{Kullback--Leibler divergence:}
The Kullback-Leibler (KL)-divergence (\cite{kullback:information}, 
also called relative entropy)
\begin{equation}
D(p|q) = \int p(z) \log\frac{p(z)}{q(z)} \rd z
\end{equation}
measures the deviation between the distribution of two random
variables with probability densities $p(z)$ and $q(z)$.  The
KL-divergence is not symmetric in its arguments (it is not a
distance).
For discrete probability distributions the interpretation is
straightforward.  The information content in the discrete probability
distribution $\bfp=\{p_i\}_{i=1}^\infty$ with $\sum_{i=1}^\infty
p_i=1$ is measured by the (information) entropy
\cite{shannon:mathematical}
\begin{equation}
H = \, -\!\sum_{i=1}^\infty p_i \log p_i  ,
\end{equation}
then the KL-divergence
\begin{equation}
D(\bfp|\bfq) = \sum_{i=1}^\infty p_i (\log p_i -\log q_i)  
\end{equation}
measures the information lost, if the probability distribution
$\bfq$ is used to approximate the true probability distribution
$\bfp$.
This characterisation carries over to the continuum.  Up to a
multiplicative constant the KL-divergence is a unique measure of
divergence (see \cite{hobson:comparison} for details).

\section{Details of the implementation}
\label{sec:implement}

We have chosen the statistical package \texttt{R}\,\cite{r:language} as
the basic tool for our computations\footnote{If you plan to use
  \texttt{python} you may consider the modules \texttt{scipy,
    statsmodels, arviz} and \texttt{pandas}, and 
    see also \cite{akeret:cosmohammer} for CosmoHammer.
    A helpful page about
  python implementations for MCMC and nested sampling is maintained by
  Matthew Pitkin
  \url{http://mattpitkin.github.io/samplers-demo/pages/samplers-samplers-everywhere}. }.
The results are presented in section\,\ref{sec:application}.  Here you
find some details about the implementation and the packages we use.
\begin{compactitem}[-]
\item We calculate the minimum $\bftheta^\starii$ of the $\chi^2$
  according to eq.\,(\ref{eq:chi2}) using the function \texttt{nls}
  from the core of \texttt{R} \cite{r:language}. The $p$--values are
  calculated using the built in $\chi^2$--distribution function.
\item For the maximum likelihood estimate $\bftheta^\stari$ we use the
  function \texttt{mle2} from the package \texttt{bbmle}
  \cite{r:bbmle}. With this $\bftheta^\stari$ we calculate the maximum
  value of the likelihood $p_f(\bfd\,|\,\bftheta^\stari)$ which we use
  in the computation of the likelihood ratio. And again we use the
  built in $\chi^2$--distribution function to calculate the $p$--value
  for the likelihood ratio test (see section\,\ref{sec:lrt}).
\item Since we only consider a one-- and a two--dimensional parameter
  space, we are able to calculate the evidence by direct numerical
  integration. We use the builtin function \texttt{integrate} and an
  adaptive multidimensional integration routine \texttt{hcubature}
  from the package~\texttt{cubature} \cite{r:cubature}.  The direct
  numerical integration gives similar results compared to the quite
  noisy and costly results obtained from nested sampling using the
  package \texttt{RNested}~\cite{r:rnested}.
  The BIC (eq.\,(\ref{eq:BIC})) is calculated using a function
  provided in the package~\texttt{bbmle} \cite{r:bbmle}.
\item For the classical information theoretic approach we first
  calculate the AIC (see eq.\,(\ref{eq:AIC})) using functions provided
  in the \texttt{bbmle} package~\cite{r:bbmle}. To go beyond this
  asymptotic result we calculate $\widehat{\eta}(f)$ according to
  eq.\,(\ref{eq:etahat}). With the bootstrap estimate
  $\widetilde{b}(f)$ of the bias $b(f)$ (see eq.\,(\ref{eq:bf})) we
  calculate the $\text{EIC}(f)$, see eq.\,(\ref{eq:EIC}).  In
  section\,\ref{sec:bootstrap} we give a detailed description of this
  bootstrap procedure due to Konishi and
  Kitagawa\,\cite{konishi:generalised}. We use 100k bootstrap samples
  to estimate $\widetilde{b}(f)$.
\item We prepare Markov chains with the function \texttt{metrop} from
  the \texttt{mcmc}~package \cite{r:mcmc}. For convergence diagnostics
  and for tuning of the sampler parameters we employ the \texttt{coda}
  package \cite{r:coda}. Using Gelman and Rubin's convergence
  diagnostic \cite{gelman:inference} we see that all our chains
  converge after at least 1000 steps, even if we start in the extreme
  points of the parameter range.
  For the $\Lambda$CDM model we build a chain with a length of 15\,Mio
  steps. We average the results over 15\,steps and use this batched
  chain for the MAP estimate and to calculate the BPIC (see next
  point). For the $w$CDM model we build a chain with a length of
  30\,Mio steps and average the results over 30\,steps.
\item We calculate the BPIC$(f)$ as described in
  section\,\ref{sec:bayesinfo}. In the $\Lambda$CDM model we estimate
  for each data point $d_i=(z_i,\mu_i)$
  \begin{equation}
    \mathbb{E}_\text{post}\left[ p_{\Lambda}(d_i\,|\,\Omega_{m})\right] \approx
    \frac{1}{L} \sum_{l=1}^L p_{\Lambda}(d_i\,|\,\Omega_{m,l}), 
  \end{equation}
  from one Markov chain. Here $\Omega_{m,l}$ is one state from the
  Markov chain and $L$ is the length of the chain.  Inserting this
  estimate of
  $\mathbb{E}_\text{post}\left[ p_{\Lambda}(d_i\,|\,\Omega_{m})\right]$
  into eq.\,(\ref{eq:hatkappa}) we obtain $\widehat{\kappa}_\Lambda$ as an
  average over all the data points. Then we rescale as in eq.\,(\ref{eq:BPIC}) 
  to obtain $\text{BPIC}_\Lambda$. We proceed similarly for the
  $\text{BPIC}_w$.
\end{compactitem}
You can download an abbreviated version of our code from
\url{https://homepages.physik.uni-muenchen.de/~Martin.Kerscher/software/modelselect/} .

\subsection{Bootstrap for $b(f)$}
\label{sec:bootstrap}

Before we describe the bootstrap procedure leading to the EIC
\cite{ishiguro:bootstrapping,konishi:generalised} we give a more
detailed definition of the average bias.
We augment the notation from Sect.~\ref{sec:information-classic} and
express the expected log likelihood of the model $f$, the data set
$\bfd=(x_i,y_i)_{i=1}^N$, and the cumulative distribution functions
$F_T$ as
\[ 
\eta(f; \bftheta^\stari(\bfd), F_T) 
 = \int \log p_f\left( d\,|\,\bftheta^\stari(\bfd)\right)\, \rd F_T(d),
\]
where $\bftheta^\stari(\bfd)$ is the best maximum likelihood parameter
obtained from the data set $\bfd$.
The average bias from eq.\,(\ref{eq:bf}) can be expressed as
\begin{equation}
\label{eq:bf_ii}
b(f) = \mathbb{E}_{F_T}\left[ \eta\left(f; \bftheta^\stari(\bfd), F_T\right) 
	- \eta\left(f; \bftheta^\stari(\bfd), F_{T,N,\bfd} \right) \right] .
\end{equation}
The dependence of the estimated parameters $\bftheta^\stari(\bfd)$ and
the empirical distribution function $F_{T,N,\bfd}$ on the data set
$\bfd$ is now explicit.
\cite{ishiguro:bootstrapping, konishi:generalised} propose a bootstrap
procedure to estimate $b(f)$.  First generate bootstrap samples
$\widetilde{\bfd}=(\tilde{x}_i,\tilde{y}_i)_{i=1}^N$ from the data by
repeatedly drawing from $\bfd$ with putting back (i.e.\ sampling from
$F_{T,N,\bfd}$).  For each of these bootstrap samples
$\widetilde{\bfd}$ we have the empirical distribution function
$F_{T,N,\widetilde{\bfd}}$.
The bootstrap estimate of $b(f)$, as given in eq.\,(\ref{eq:bf_ii}),
is then\footnote{Please watch where we write $\widetilde{\bfd}$ or
  $\bfd$.}  
\begin{align}
  \widetilde{b}(f)
  & = \widetilde{\mathbb{E}}\left[ 
    \eta\left(f; \bftheta^\stari(\widetilde{\bfd}), F_{T,N,\bfd}\right) 
    - \eta\left(f; \bftheta^\stari(\widetilde{\bfd}), F_{T,N,\widetilde{\bfd}} \right)
    \right] \nonumber \\
  & = \widetilde{\mathbb{E}}\left[
    \frac{1}{N}\sum_{i=1}^N
       \log p_f\left(d_i\,|\,\bftheta^\stari(\widetilde{\bfd})\right) 
    - \frac{1}{N}\sum_{j=1}^N
       \log p_f\left(\widetilde{d}_j\,|\,\bftheta^\stari(\widetilde{\bfd})\right)
    \right] .
\end{align}
The expectation
$\widetilde{\mathbb{E}}[\cdot]\equiv\mathbb{E}_{F_{T,N,\bfd}}[\cdot]$
is over samples $\widetilde{\bfd}$ drawn from $F_{T,N,\bfd}$.  Using
$M$ such bootstrap samples
$\widetilde{\bfd}^\alpha, \alpha=1,\ldots,M$ we can estimate
$\widetilde{b}\left(f\right)$ by
\begin{equation}
  \widetilde{b}\left(f\right) \approx \frac{1}{MN}\sum_{\alpha=1}^M \sum_{i=1}^N
  \log\left(\frac{p_f\left(d_i\,|\,\bftheta^\stari(\widetilde{\bfd}^\alpha)\right)}
                {p_f\left(\widetilde{d}_i^\alpha\,|\,\bftheta^\stari(\widetilde{\bfd}^\alpha)\right)}
                \right) .
\end{equation}
Depending on the estimation procedure for
$\bftheta^\stari(\widetilde{\bfd})$, such a bootstrap procedure can be
time consuming.  Konishi \& Kitagawa \cite{konishi:generalised} show
that $\widetilde{b}(f)$ is approximating $b(f)$ for large $N$.
Furthermore they propose a variance reduction scheme for this
bootstrap procedure.

\section{More on the data analysis}
\label{sec:errors}
The analysis of the SN\,Ia data in section\,\ref{sec:application}
serves as an illustrative example for the methods of model
selection. To keep things simple we employ some approximations,
specifically we assume a diagonal covariance matrix in the likelihood
and also assume that the variances are independent from the
cosmological model.  Below we will try to give justice to the more
complex situation.

The distance moduli $\mu_i$ of the SN\,Ia are calculated with a (semi)
empirical relation from the observed light curve of the supernova
explosion. Several parameters enter this relation (see
\cite{kowalski:improved} for details). In our analysis we use the
$\mu_i$ provided in the Union\,2.1 compilation which have been
calculated with the best fit parameters. Such a uniform fitting
introduces covariances between the $\mu_i$. They have been estimated
and the Union\,2.1 compilation comes with a non diagonal covariance
matrix (see \cite{suzuki:hubble} and
\url{http://supernova.lbl.gov/Union/}).  In a full analysis we would
have to include these covariances in the likelihood (compare
eq.\,(\ref{eq:gauss-likelihood})). Moreover in a full Bayesian
analysis we would include these fitting parameters as independent
parameters and then later marginalise (compare
\cite{gelman:philosophy}).
Further error sources are photometric zero points, contamination,
evolution, Malmquist bias, K-corrections, gravitational lensing,
peculiar velocities, etc. (see \cite{kowalski:improved}). They all
contribute to the (co-)variances and have been estimated in the
Union\,2.1 compilation \cite{suzuki:hubble}.

Some of these contributions to the error budget also depend on the
cosmological model.  For example the magnification and demagnification
of high redshift supernovae by gravitational lensing depends on the
structure growth, which again depends on the cosmological parameters.
This lensing contribution can be estimated for some of the supernovae
individually \cite{suzuki:hubble} but often this lensing error is
estimated in a statistical sense only \cite{holz:safety}.  Then
probably a self consistent treatment will be necessary if one aims for
higher precision.
Also anisotropies and inhomogeneities in the matter distribution
influence the obeservations \cite{szapudi:detection}.  A careful
determination of the errors will be necessary if one compares with
inhomogeneous models
\cite{enqvist:effect,larena:testing,lawrence:apparent}.  Not only the
distance modulus redshift relation but also the errors depend on the
adopted models and have to be quantified.

To get a rough idea how these additional uncertainties influence our
results we apply a uniform scaling factor $C$ to the $\sigma_{\mu,i}$
and then repeat the analysis from section\,\ref{sec:application}.  As
can be seen from table\,\ref{tab:variances} the values of the relevant
quantities change, but compared to the dispersion estimates from the
mock samples shown in table\,\ref{table:dispersion}, the observed
differences between the $\Lambda$CDM and the $w$CDM model remain
small.

\begin{table}
  \caption{The values of the relevant quantities for model selection 
    obtained from the Union\,2.1 sample after scaling the 
    uncertainties $\sigma_{\mu,i}$ by a factor $C$.}
  \label{tab:variances}
\begin{center}
\begin{tabular}{|l|c|c|c|c|c|c|}
  \hline
  \diagbox{}{ {\phantom{xxx}C}}  & 0.5 & 0.8  & \textbf{1.0} & 1.2 & 1.5 & 2\\
  \hline
  \hline
  $p_\Lambda$                &  0     & 0     & 0.684 & 1 & 1 & 1\\
  $p_w$                    &  0     & 0      & 0.673 & 1 & 1 & 1\\
  \hline
  $p$  l-ratio test        &  0.951 & 0.969  & 0.975 & 0.980 & 0.984 & 0.988\\
  \hline
  $B_{\Lambda w}$             & 23.3 & 6.82 & 5.45 & 4.54 & 3.66 & 2.41 \\
  \hline
  $\text{BIC}_\Lambda$       &  651.5 & -173.7 & -231.1 & -191 & -73.1 & 151.3\\
  $\text{BIC}_w$            &  657.9 & -167.4 & -224.8 & -185 & -66.8 & 157.6\\
  \hline
  $\text{AIC}_\Lambda$        &  647.1 & -178.1 & -235.5 & -195.8 & -77.5 & 146.9\\
  $\text{AIC}_w$             &  649.1 & -176.1 & -233.5 & -193.8 & -75.5 & 148.9\\
  \hline
  $\text{EIC}_\Lambda$  & 641.5 & -181.4 & -239.2 & -198.2 & -79.7 & 353.2 \\
  $\text{EIC}_w$       &  632.1 & -185.4 & -241.0 & -200.3 & -81.2 & 352.4 \\
  \hline
  $\text{BPIC}_\Lambda$  & 643.9 & -180.2 & -237.5 & -197.7 & -79.3 & 145.1\\
  $\text{BPIC}_w$       & 641.9  & -180.5 & -237.3 & -197.4 & -78.9 & 145.6 \\
  \hline
\end{tabular}
\end{center}
\end{table}

\end{appendix}

\footnotesize
\bibliography{my}


\end{document}